\documentclass[fleqn]{2023SCGE}
\usepackage{graphicx}
\usepackage{float}

\begin{document}

\ensubject{subject}

\ArticleType{Article}
\SpecialTopic{SPECIAL TOPIC: }
\Year{}
\Month{}
\Vol{}
\No{}
\DOI{??}
\ArtNo{000000}
\ReceiveDate{}
\AcceptDate{}

\title{Forecast on $f(R)$ Gravity with HI 21-cm Intensity Mapping Surveys}{Forecast on $f(R)$ Gravity with HI 21-cm Intensity Mapping Surveys}

\author[1]{Yanling Song}{}%
\author[1]{Yu Sang}{sangyu@yzu.edu.cn}
\author[2]{Linfeng Xiao}{}
\author[1,3]{Bin Wang}{}

\AuthorMark{Y. Song}

\AuthorCitation{Yanling Song, Yu Sang, Linfeng Xiao, Bin Wang}
\address[1]{Center for Gravitation and Cosmology, College of Physical Science and Technology, Yangzhou University, Yangzhou, 225009, China}
\address[2]{Department of Physics and Astronomy, Sejong University, Seoul, 143-747, Korea}
\address[3]{School of Aeronautics and Astronautics, Shanghai Jiao Tong University, Shanghai 200240, China}


\abstract{Modified gravity theories offer a well-motivated extension of General Relativity and provide a possible explanation for the late-time accelerated expansion of the Universe. Among them, $f(R)$ gravity represents a minimal and theoretically appealing class, characterized by the Compton wavelength parameter $B_0$, which quantifies deviations from General Relativity. In this work, we explore the capability of future neutral hydrogen (HI) 21-cm intensity mapping (IM) observations to constrain $f(R)$ gravity at low redshifts. 
We perform Fisher-matrix forecasts for $B_0$ and standard cosmological parameters using upcoming 21-cm IM experiments, including BINGO and SKA1-MID (Band 1 and Band 2), both individually and in combination with Planck cosmic microwave background (CMB) priors.  
For the phenomenological HI treatment, we obtain $\sigma(B_0)=3.73\times10^{-6}$, $5.98\times10^{-6}$, and $6.78\times10^{-8}$ for BINGO, SKA1-MID Band 1, and Band 2, respectively, and the tightest bound is improved to $3.95\times10^{-8}$ when Planck priors are included. We further consider a redshift-dependent HI model with fixed $\Omega_{\rm HI}(z)$ and $b_{\rm HI}(z)$, and find that the constraints remain of similar order, confirming the robustness of the results. In both cases, SKA1-MID Band 2 provides the strongest sensitivity, while foreground residuals degrade the constraints but do not change the main conclusion that future HI intensity mapping, especially combined with CMB data, can provide stringent tests of General Relativity on cosmological scales.

}
 


\maketitle


\begin{multicols}{2}
\section{Introduction}\label{section1}
Observations of Type Ia supernovae indicate that the Universe is undergoing accelerated expansion \cite{perlmutter1999measurements}. The $\Lambda$CDM cosmological model is the simplest explanation for the accelerated expansion and provides a good fit to most cosmological observations. However, the $\Lambda$CDM cosmological model is challenged by the inconsistent cosmological parameters inferred from different measurements. The Hubble parameter derived from cosmic microwave background (CMB) 
measurements is inconsistent with that from 
\Authorfootnote
the SH0ES (Supernovae and $H_0$ for the EoS of DE) at the 5$\sigma$ confidence level, which is called Hubble tension \cite{Riess_2022, aghanim2020planck}. Another tension is of $\sigma_8$, the mass variance on a scale of 8$\rm{Mpc}/h$, which comes from the 3$\sigma$ discrepancy between the measurements of the CMB lensing and cosmic shear surveys \cite{Asgari_2021, Doux_2022, aghanim2020planck}. These observational tensions motivate the studies of modification of General Relativity (GR) on cosmological scales. Among the alternative theories, $f(R)$ gravity has been widely studied due to its simple form and clear physical descriptions. It is a modified gravity theory that adds a general function $f(R)$ of the Einstein-Hilbert action $R$, introducing a scalar degree of freedom called the scalaron that regulates gravitational interactions on cosmological scales \cite{starobinsky1980new, De_Felice_2010, Katsuragawa_2018,
Ishak:2024jhs,Artis:2024eco,Vogt:2024SPTfR, Bai:2024hpw}.

A key parameter in $f(R)$ gravity is $B_0$, which represents the present-day value of the Compton wavelength of the scalaron expressed in units of the Hubble length \cite{De_Felice_2010, Sotiriou_2010}. This parameter quantifies the deviation of $f(R)$ gravity from GR. When $B_0 = 0$,  the $\Lambda$CDM model is recovered, whereas non-zero value of $B_0$ indicates  gravitational modifications of the expansion history of the universe and the formation of large-scale structure (LSS) \cite{NOJIRI_2007}. Due to the limited sensitivity of observational probes, current cosmological observational data impose constraints on $B_0$ at order of $10^{-4}$. Using CMB alone, the constraint is $B_0 < 3.37$ for WMAP9 and $B_0 <0.18$ for Planck at 95\% confidence level (C.L.) \cite{Hu2013}. LSS probes based on galaxy clustering, cluster abundance, growth rates and galaxy power spectra  place strong constraints on $B_0$. Inclusion of data from cluster abundance to the measurement from supernovae, acoustic oscillation (BAO), Hubble constant and CMB, is largely improving the constraints to $B_0 < 1.1 \times 10^{-3}$ at the 95\% C.L. \cite{Lombriser2011}. The combination of Planck CMB data and BAO data yields tight constraints on $B_0$, with $B_0 < 0.006$ at the 95\% C.L. for $w=-1$, and $B_0 < 0.0045$ (95\% C.L.) when $w \neq -1$ \cite{Battye2018}. WiggleZ power spectrum in addition with CMB data from Planck give $\log_{10} B_0 < -4.07$ at 95\% C.L. \cite{Dossett2014}. And cluster abundance plus CMB and BAO yields a constraints of $\log_{10} B_0 < -3.68$ at the 95\% C.L. \cite{Cataneo2015}. Scale-dependent growth rates of structure combined with  CMB and BAO give $\log_{10} B_0 < -4.1$ at 95\% C.L. \cite{Li2015}. In addition to constraining the parameter $B_0$, there are some other works focusing on testing $f(R)$ gravity from other aspects \cite{Benisty:2023qcv, Liu:2016xes, Peirone:2016wca, Jana:2018djs, Liu:2021weo, Kou:2023gyc, Vogt:2024pws}. 

In addition to current high-precision measurements from CMB and galaxy survey, neutral hydrogen (HI) 21-cm intensity mapping (IM) has emerged as a powerful probe of cosmology and modified gravity in the radio waveband \cite{Hall2013,Battye2004,Madau1997}. Unlike traditional galaxy surveys, 21-cm IM does not require the resolution of individual galaxies. The primary objective of HI IM is to map the integrated 21-cm radiation intensity from numerous unresolved galaxies across a given redshift range \cite{Furlanetto2016, mirocha2019astrophysics21cmbackground}. This signal encodes information on matter density perturbations, redshift-space distortions (RSD), and gravitational potentials, making it sensitive to signatures of modified gravity  \cite{Hall2013}. In this work, we focus on constraining $f(R)$ gravity using 21-cm IM data from BINGO and SKA. BINGO is a single-dish telescope in Brazil, covering 3000 deg$^2$ at redshifts $0.13 \leq z \leq 0.45$ (980–1260 MHz) with an angular resolution of 40 arcmin \cite{Bingo1,Bingo2,Bingo3,Bingo4,Bingo5,Bingo6,Bingo7,dosSantos:2023ipw,Zhang:2024bar,Motta:2026til}. SKA1-MID is a dish array in South Africa, operating in two bands: Band 1 (350–1050 MHz, $0.36 \leq z \leq 3.0$) covering 20000 deg$^2$ and Band 2 (950–1410 MHz, $0.01 \leq z \leq 0.49$) covering 5000 deg$^2$, with significantly lower system temperatures (15 K for Band 2) than BINGO (70 K)\cite{SKA_redbook2020}. These experiments can further reduce noise and are expected to place tight constraints on $f(R)$ gravity. 

In this work, we forecast the constraints on $B_0$ using 21-cm IM  from BINGO and SKA1-MID, combined with CMB observations. In Section~\ref{sec:2}, we introduce the  $f(R)$ gravity, with its background and  perturbation evolutions.  In Section~\ref{sec:3}, we describe the redshifted 21-cm HI signal in the framework of $f(R)$ gravity, including the brightness temperature evolutions at the background and perturbation level. We also include the angular power spectrum of the HI signal.  In Section \ref{sec:4}, we present the thermal and shot noise for each experiment, and using the Fisher matrix method to forecast constraints on $B_0$ and key cosmological parameters. In Sections \ref{sec:results} and \ref{sec:results2}, we present the forecast results for the phenomenological and redshift-dependent HI models, respectively, in Section \ref{sec:conclusions} we present the conclusions and discussions. 

\section{$f(R)$ gravity models}
\label{sec:2}
We focus on the following action in Jordan frame given by 
\begin{equation}
    S=\frac{1}{16\pi G}\int d^4x\sqrt{-g}(R+f(R))+S_m, 
\end{equation}
where $g$ is the determinant of the metric, $R$ is the Ricci scalar and $f(R)$ is an arbitrary function of $R$, $S_m$ is the action for matter fields. 
Taking the variation of the above action with respect to the metric, one get the modified Einstein equations \cite{Song2007,Hu2007,Pogosian2008}:
\begin{equation}
    G_{\alpha\beta}+f_R R_{\alpha\beta}- (\frac{f}{2}-\Box f_R)g_{\alpha\beta} - \nabla_\alpha\nabla_\beta f_R = \kappa^2T_ {\alpha\beta},
\end{equation}
where $f_R=df(R)/dR$, $\kappa^2=8\pi G$.
When giving the flat FLRW metrc $ds^2=a^2(\tau)(d\tau^2-d \textbf{\textit{x}}^2)$, one can get the background evolutions \cite{Song2007,Pogosian2008,Hojjati2012}:
\begin{eqnarray}
    (1+f_R)\mathcal H^2+\frac{a^2}{6}f-\frac{\ddot a}{a}f_R+ \mathcal H\dot f_R=\frac{\kappa^2}{3}a^2\rho \\
    \frac{\ddot a}{a}- (1+f_R)\mathcal H^2+\frac{a^2}{6}f+\frac{1}{2}\ddot f_R=-\frac{\kappa^2}{6}a^2(\rho+3P)
\end{eqnarray}
where $\mathcal H=\dot a/a$,  and an overdot denotes differentiation with respect to conformal time. Since $f$ is an arbitrary function of $R$, the background evolutions can be tunned to fit any kinds of $\Lambda$CDM or non-$\Lambda$CDM histories.  And we define the Compton wavelength of the scalaron in Hubble units \cite{Song2007}:
\begin{equation}
    B=\frac{f_{RR}}{1+f_R}\frac{\mathcal H \dot R}{\dot{\mathcal H}-\mathcal H^2}.
\end{equation}
One can find a family of $f$ forms  for the  fitting of a given expansion history through $B$'s present value $B_0$, or equivalently, through the boundary condition $f_R^0$. In $\Lambda$CDM case, $B_0$ can be related to $f_R^0$ approximately by $B_0\approx -6f_R^0$ \cite{Raveri2014}.

For perturbations, we consider scalar perturbations in the Newtonian gauge, with the line element given by
\begin{equation}
    ds^2=a^2[(1+2\Psi)d\tau^2-(1-2\Phi)d\textbf{\textit{x}}^2],
\label{eq:metric}
\end{equation}
and the perturbed energy-momentum tensor is
\begin{eqnarray}
\label{eq:energe-momentum}
    T_0^0&=&-\rho(1+\delta),\nonumber\\
    T_i^0&=&-(\rho+P)v_i, \\
    T_i^j&=&(P+\delta P) \delta_i^j, \nonumber
\end{eqnarray}
where $\delta=\delta\rho/\rho$ is the density contrast, $v$ is the  velocity, $\delta P$ the pressure perturbation, $\delta_i^j=0$ for $i\neq j$ and $\delta_i^j=1$ for $i=j$.  In the Jordan frame, the metric $g_{\mu\nu}$ is minimally coupled to matter,
so the energy-momentum is conserved. And the continuity and Euler equations retain their forms as in the standard GR, in Fourier space, they become \cite{Hall2013}
\begin{eqnarray}
    \dot\delta+kv-3\dot\Phi&=&0,\\
    \dot v+\mathcal Hv-k\Psi&=&0.
    \label{eq:Euler}
\end{eqnarray}
Then one can see that the differences under modified gravities for $\delta, v$ is decipted in $\Phi, \Psi$.   For the full set of perturbation equations, one can submit equations (\ref{eq:metric}) and (\ref{eq:energe-momentum}) to solve for the modified  Einstein 
equations. One can refer to Ref. \cite{Pogosian2008} for the explicit expression of these equations. Here we want to focus on the effective anisotropy and the modified Poisson equations that generally describe the deflections of modified gravities from GR. As in MGCAMB \cite{Wang2023}, the modification to Poisson and anisotropy equations are encoded in two functions $\mu(a,k)$ and $\gamma(a,k)$ defined by
\begin{eqnarray}
    k^2\Psi&=&-\frac{a^2}{2M_{pl}^2}\mu(a,k)\rho\Delta,  \\
    \frac{\Phi}{\Psi}&=&\gamma(a,k),
\end{eqnarray}
where $M_{pl}$ is the Planck mass, and $\rho\Delta=\rho\delta+3\frac{aH}{k}(\rho+P)v$ is the comoving density perturbation. On quasi-static scales the $B_0$
parameterization of $f(R)$ gravity can be parameterized by \cite{Hojjati2011, Giannantonio2010,Bingo7}
\begin{eqnarray}
    \mu(a,k)&=&\frac{1}{1-1.4\times10^{-8}|\lambda/{\rm Mpc}|^2a^3} \frac{1+\frac{4}{3}\lambda^2k^2a^4}{1+\lambda^2k^2a^4}, \\ 
    \gamma(a,k)&=&\frac{1+\frac{2}{3}\lambda^2k^2a^4}{1+\frac{4}{3}\lambda^2k^2a^4}, 
\end{eqnarray}
where $B_0=2H_0^2\lambda^2$.

\section{21-cm angular power spectra }
\label{sec:3}

In this section we present the brightness temperature fluctuations of the redshifted 21-cm signal, emitted from the hyperfine transition between the excited triplet states and the singlet states of neutral hydrogen atoms. It is fine to neglecte the finite line width of the emission on large scales. And the stimulated emission and absorption can be neglected at low redshift. The 21-cm brightness temperature $T_{\rm b}$ is \cite{Hall2013}: 
\begin{equation}
     T_{\rm b}(z,\textbf{\textit{n}})=\frac{3}{32\pi}\frac{h_{\rm p}^3n_{{\rm HI}}A_{10}}{k_{\rm B}E_{21}}|\frac{{\rm d}\lambda}{{\rm d}z}|.
\label{eq:brightness1}
\end{equation}
Here $h_{\rm p}$ is the Planck's constant, $A_{10}\approx 2.869\times10^{-15}{\rm s}^{-1}$ is the spontaneous emission coefficient, $k_{\rm B}$ is Boltzmann’s constant, and $E_{21}=5.88\mu{\rm eV}$ is the rest-frame energy of a 21-cm photon. $\textbf{\textit{n}}$ is the unit vector along the line of sight, $n_{{\rm HI}}$ is the  rest-frame number density of neutral hydrogen atoms,  $\lambda$ is an affine parameter for the light path.
Since the distribution of HI is a good tracer of the large scale structures of our Universe, it is quite natural to expand equation (\ref{eq:brightness1}) to its background level and perturbation level, as have been down in the analysis of the large scale structure.

If we first exclude the contribution from perturbations, $|{\rm d}z/{\rm d}\lambda|=(1+z)H(z)E_{21}$, then the background brightness temperature is 
\begin{equation}
    \bar T_{\rm b}(z)=\frac{3(h_{\rm p}c)^3\bar n_{{\rm HI}}A_{10}}{32\pi k_{\rm B}E_{21}^2(1+z)H(z)}=0.188 {\rm K} h\Omega_{{\rm HI}}(z)\frac{(1+z)^2}{E(z)}.
\end{equation}
Here $H(z)$ is the Hubble parameter, $E(z)=H(z)/H_0$, $H_0$ is the Hubble constant. $\Omega_{{\rm HI}}(z)$ is the comoving mass density of HI in units of the current critical density. 

When considering perturbations, 
\begin{align}
n_{\rm HI}(z,\textbf{\textit{n}}) 
&= \bar{n}_{\rm HI}(z)\big(1+\delta_n(z,\textbf{\textit{n}})\big) 
= \bar{n}_{\rm HI}(\bar{\tau}_z+\delta\tau)\big(1+\delta_n(z,\textbf{\textit{n}})\big) \nonumber\\
&= \bar{n}_{\rm HI}(\bar{\tau}_z)\left( 1 + \frac{\dot{\bar{n}}_{\rm HI}}{\bar{n}_{\rm HI}}\delta\tau + \delta_n \right), \\[6pt]
\left| \frac{{\rm d}\lambda}{{\rm d}z} \right|(z,\textbf{\textit{n}})
&= \frac{a(\bar{\tau}_z)}{\mathcal{H}(\bar{\tau}_z) E_{21}(1+z)} \Bigg[
1 - \left( \frac{\dot{\mathcal{H}}}{\mathcal{H}} - \mathcal{H} \right)\delta\tau
- \frac{1}{\mathcal{H}}\frac{{\rm d}\Psi}{{\rm d}\tau} \nonumber\\
&\qquad + \frac{1}{\mathcal{H}}(\dot{\Phi}+\dot{\Psi})
+ \frac{1}{\mathcal{H}}\textbf{\textit{n}}\cdot\frac{{\rm d}\textbf{\textit{v}}}{{\rm d}\eta}
+ \Psi + \textbf{\textit{n}}\cdot\textbf{\textit{v}} \Bigg],
\end{align}
so the perturbation to the brightness temperature is
\begin{align}
\Delta_{T_{\rm b}}(z, \textbf{\textit{n}})
= \frac{\delta T_{\rm b}(z,\textbf{\textit{n}})}{\bar T_{\rm b}(z)} 
= \delta_n+\frac{\dot{\bar n}_{\rm HI}}{n_{\rm HI}}\delta\tau-\left(\frac{\dot{\mathcal H}}{\mathcal H}-\mathcal H\right)\delta\tau \nonumber\\
   -\frac{1}{\mathcal H}\frac{{\rm d}\Psi}{{\rm d}\tau}  
  + \frac{1}{\mathcal H} (\dot\Phi+\dot\Psi) 
 +\frac{1}{\mathcal H} \textbf{\textit{n}}\cdot \frac{{\rm d}\textbf{\textit{v}}}{{\rm d}\tau} +\textbf{\textit{n}}\cdot\textbf{\textit{v}}+\Psi.
\end{align}
The above equation is valid for all metric theories of gravity, including $f(R)$ gravity, one can refer to \cite{Hall2013} for the details of the inferring process. Using the Euler equation (\ref{eq:Euler}) but in configuration space $\dot{\textbf{\textit{v}}}+\mathcal{H}\textbf{\textit{v}}+\nabla\Psi=0$, it becomes
\begin{equation}
    \Delta_{T_{\rm b}}(z,\textbf{\textit{n}})=\delta_n-\frac{1}{\mathcal{H}}\textbf{\textit{n}}\cdot[(\textbf{\textit{n}}\cdot\nabla)\textbf{\textit{v}}]+(\frac{{\rm d}{\rm ln}(a^3\bar n_{{\rm HI}})}{{\rm d}\tau}-\frac{\dot{\mathcal{H}}}{\mathcal{H}} -2\mathcal{H})\delta\tau +\frac{1}{\mathcal{H}}\dot\Phi+\Psi.
\label{eq:brightness2}
\end{equation}
In this equation, the first term is the usual density term, and the second is the RSD term. The other term also have their physical meanings and origins, for examlpe, the integrated Sachs-Wolfe (ISW) effect and Doppler shift. The first two terms are dominant over other terms \cite{Hall2013,Bingo7,Xiao2021}, so here we just consider the matter density term and the RSD term in our later analysis. Since the distribution of HI gas follows that of the discrete sources in which the neutral gas reside in, there is some bias between the HI distribution and the underlying matter distribution. And we follow the assumption that the bias is scale-independent at the linear order. During the period of matter domination, the comoving gauge coincides with the synchronous gauge, then we have \cite{Hall2013,Xiao2021}
\begin{equation}
    \delta_n=b_{{\rm HI}}\delta_{\rm m}^{{\rm syn}}+ \Big ( \frac{{\rm d}{\rm ln}(a^3\bar n_{{\rm HI}})}{{\rm d}\tau} - 3\mathcal H \Big )\frac{v_{\rm m}}{k},
\end{equation}
where $b_{{\rm HI}}$ is the scale-independent bias, $v_{\rm m}$ is the Newtonian-gauge matter velocity and $\delta_{\rm m}^{{\rm syn}}$ is the matter overdensity in the synchronous gauge.

Due to the projection of the three dimensional HI distribution into the two dimensional sphere, for a fixed redshift, one can expand the brightness temperature in spherical harmonics, that is:
\begin{equation}
    \Delta_{T_{\rm b}}(z,\textbf{\textit{n}})=\sum_{lm}\Delta_{T_{\rm b},lm}(z)Y_{lm}(\textbf{\textit{n}}).
\end{equation}
Then express the perturbation coefficients $\Delta_{T_{\rm b},lm}(z)$ using the Fourier transforms of the perturbations, we have 
\begin{equation}
    \Delta_{T_{\rm b},lm}(z) = 4\pi i^l\int \frac{{\rm d}^3\textbf{\textit{k}}}{(2\pi)^{3/2}}\Delta_{{\rm b},l}(z,\textbf{\textit{k}})Y_{lm}^*(\textbf{\textit{{k}}}).
\end{equation}
Following equation (\ref{eq:brightness2}), the $l$th multipole moment of $\Delta_{T_{\rm b}}$ is
\begin{align}
    \Delta_{T_{\rm b},l}(z,\textbf{\textit{k}})=\delta_nj_l(k\chi)+\frac{kv}{\mathcal H}j_l^{\prime\prime}(k\chi)+(\frac{1}{\mathcal H}\dot\Phi+\Psi)j_l(k\chi)\nonumber\\ 
    -\Big (\frac{1}{\mathcal H}\frac{{\rm d}{\rm ln}(a^3\bar n_{{\rm HI}})}{{\rm d}\tau}-\frac{\mathcal{\dot H}}{\mathcal H^2} -2    \Big)\Big[ \Psi j_l(k\chi)+vj_l^{\prime}(k\chi) \nonumber\\
    +\int_0^\chi (\dot\Psi+\dot\Phi)j_l(k\chi\prime) {\rm d}\chi\prime   \Big ].
\end{align}
where $\chi$ is the comoving distance to redshift $z$, $j_l(k\chi)$ is the spherical Bessel Function. The prime in $j_l^{\prime}(k\chi),j_l^{\prime\prime}(k\chi)$ is the derivative with respect to $k\chi$. We then integrate over a redshift (frequency) normalized window function centered at redshift $z$
\begin{equation}
    W(z)= \left\{ 
\begin{array}{cc}
    {1}/{\Delta z}, & (z-{\Delta z}/{2}\leq z \leq z+ {\Delta z}/{2}) \\
    0, & {\rm otherwise}
\end{array}
\right .
\end{equation}
to get
\begin{equation}
    \Delta_{T_{\rm b},l}^W(\textbf{\textit{k}})=\int_0^\infty dzW(z)\Delta_{T_{\rm b},l}(z,\textbf{\textit{k}}).
\end{equation}
Then the angular power spectrum between redshift windows is
\begin{equation}
    C_l^{WW \prime}=4\pi \int {\rm d}{\rm ln}k\mathcal P_{\mathcal R}(k)\Delta_{T_{\rm b},l}^W(k)\Delta_{T_{\rm b},l}^{W \prime}(k).
\end{equation}
Here $\mathcal P_{\mathcal R}(k)$ is the dimensionless power spectrum of the primordial curvature perturbation $\mathcal R$, and we define $\Delta_{T_{\rm b},l}^W(k)\equiv \Delta_{T_{\rm b},l}^W(\textbf{\textit{k}})/\mathcal P_{\mathcal R}(\textbf{\textit{k}}) $.

\begin{figure}[H]
     \centering
     \includegraphics[width=1.\linewidth]{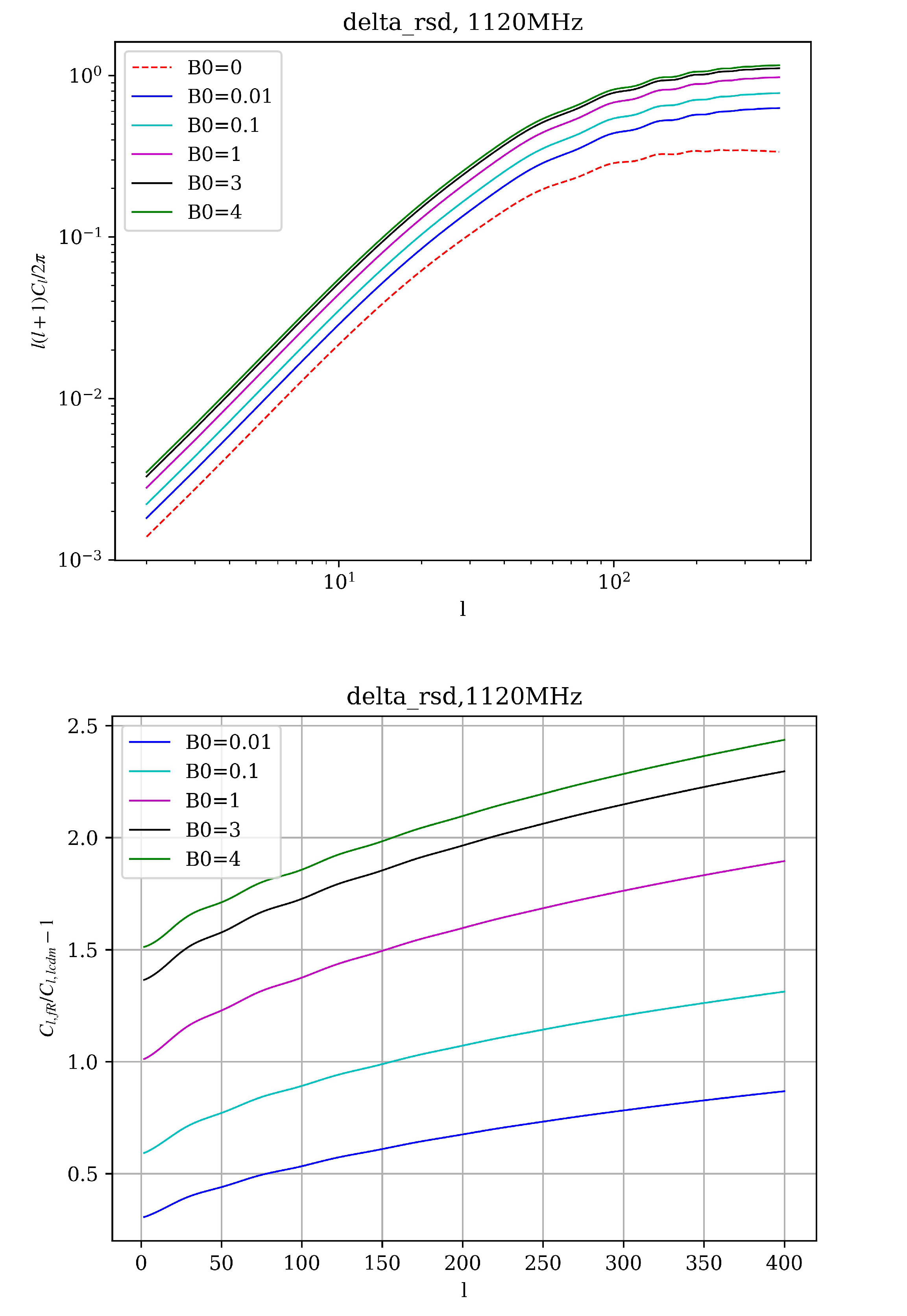}
     \caption{Angular power spectra of the HI 21-cm signal at redshift $z=0.27$ with a channel bandwidth of $10\,\mathrm{MHz}$. Top panel: auto-spectra for different values of the $f(R)$ parameter $B_0$, where $B_0=0$ corresponds to the $\Lambda$CDM case. Bottom panel: fractional deviations of the $f(R)$ spectra relative to $\Lambda$CDM. Larger values of $B_0$ lead to stronger departures from $\Lambda$CDM, especially on small angular scales. It is worth noting that this is the power spectra from theoretical calculation but not related to the experiment.  In our later analysis, we use the multipole range in $\ell = 2-400$. The maximum multipole given by the dish resolution, is $l_{\max}\gtrsim 700$ (see Ref.~\cite{Pinheiro:2026mcm} ) for this case, and therefore the angular power at $l > 400$ is not shown explicitly.   
     }   
     \label{fig: cl_fR}
\end{figure}

In Figure \ref{fig: cl_fR}, we plot the  auto-spectra for $f(R)$ gravity at redshift $z=0.27$ with the channel bandwidth being  10 ${\rm MHz}$, with just the first two terms in equation (\ref{eq:brightness2}). In the top panel, we plot the auto-spectra for $f(R)$ gravity at $B_0=0,0.01,0.1,1,3,4$. $B_0=0$ is regarded as the base case,  which is the same as $\Lambda$CDM. One can see that as the $B_0$ becomes larger, the auto-spectra become larger, too. In the bottom panel, we plot the deflections of auto-spectra for $f(R)$ gravity with respect to $\Lambda$CDM.  Similarly, the $B_0$ is larger,  the differences of auto-spectra between $f(R)$ gravity and  $\Lambda$CDM are greater.

\section{Noises and the Fisher matrix}
\label{sec:4}

To have a forecast for the constraints of the cosmological parameters from the upcoming 21-cm IM experiments, we need not only the angular power spectrum from the HI 21-cm signal, but also that from  the noise.  In our analysis, we assume the BINGO and SKA surveys as representative IM experiments.  

\subsection{surveys}

\begin{table}[H]
\footnotesize 
\tabcolsep 4pt 
\begin{tabular}{l|c c c}
\toprule
\hline
      & Bingo  & SKA Band 1 &  SKA Band 2  \\
\hline
    redshift range & [0.13,0.45] &     [0.36,3]  &  [0.01,0.49]   \\
    frequency range(MHz) & [980,1260] &   [350,1050]  &[950,1410]  \\
    Number of dishs &  1   &  197   &  197    \\
    Number of beams (dual pol.) & $28\times 2$ & $1\times 2$ & $1\times 2$  \\
    Number of channels & 28  & 70    &   46    \\
    Channel bandwidth ($\delta \nu$, MHz) & 10 & 10 & 10  \\
    Resolution ($\theta_{{\rm FWHM}}$, arcmin) & 40 & 106 & 63  \\
    System temperature ($T_{{\rm sys}}, \rm K$)  & 70  & Eq.~(\ref{eq:system temperature})    & 15   \\
    Survey time ($t_{{\rm obs}}$, yr) & 1 & 1 & 1  \\
    Sky coverage ($\Omega_{{\rm sur}}, {\rm deg}^2$)   & 3000  & 20000 & 5000 \\
\hline
\bottomrule
\end{tabular}
 \caption{Survey specifications adopted in this work for BINGO and SKA1-MID (Band~1 and Band~2). These parameters are used to compute the instrumental noise and forecast the cosmological constraints.}
\label{tab:surveys}
\end{table}

BINGO is a single-dish, drift-scan radio telescope currently under construction in the northeast of Brazil, conceived as a pathfinder for low-redshift 21-cm intensity mapping and the first-generation radio detection of BAO \cite{Bingo1}.
The baseline Phase~1 concept employs a dual-reflector optical system with two $\sim 40$\,m-class dishes in a crossed-Dragone configuration, illuminating a focal plane populated by $28$ feed horns \cite{Bingo3}. Each horn is coupled to a receiver chain providing dual-polarization measurements, often implemented as two circular polarizations. 

BINGO is designed to operate in the frequency interval $980~\mathrm{MHz} \le \nu \le 1260~\mathrm{MHz}$, which corresponds to the 21-cm rest frequency $\nu_{21}=1420$\,MHz mapped into the redshift range $ 0.127 \lesssim z \lesssim 0.449 $, thereby targeting the low-redshift regime where BAO measurements place strong leverage on late-time cosmology and dark-sector phenomenology \cite{Bingo1,Bingo7}. The survey strategy is optimized to minimize instrumental complexity and time-variable systematics by adopting a transit mode with no moving dishes \cite{Bingo2}. With the primary mirror pointing at approximately $\delta\simeq -15^\circ$, BINGO scans an instantaneous declination strip of about $14.75^\circ$ and is designed to cover $\sim 13\%$ of the sky over nominal operation \cite{Bingo2}.  To avoid nonlinear effects, we assume the channel bandwidth to be 10 MHz. BINGO has an angular resolution of $\theta_{{\rm FWHM}}=40$ arcmin, with the  full-width half-maximum (FWHM) beam resolution $\theta_{{\rm FWHM}}$ defined by $ \theta_{{\rm FWHM}}= 1.2 \lambda_{{\rm med}}/D_{{\rm dish}}$, where $D_{{\rm dish}}=34{\rm m}$ is the illuminated aperture, and $\lambda_{{\rm med}}=c/\nu_{{\rm med}}$ is the wavelength at the medium frequency $\nu_{{\rm med}}$ of the entire range \cite{Bingo1,Bingo2,Bingo3}. And $\theta_{{\rm FWHM}}=40$ arcmin  corresponds to the angular resolution of BINGO at 1 GHz.

The SKA is an international effort to construct a next-generation radio observatory, which will ultimately achieve a total collecting area of order one square kilometre, providing unprecedented sensitivity for cosmological and astrophysical studies \cite{SKA_redbook2020}. The SKA will be implemented in two major phases, referred to as SKA1 and SKA2. While SKA1 is currently under construction, the detailed configuration of SKA2 has not yet been finalized and will be defined at a later stage. Phase~1 of the project consists of two complementary instruments: SKA1-MID and SKA1-LOW. SKA1-MID is a dish-based array located in the Northern Cape province of South Africa, whereas SKA1-LOW is a low-frequency aperture array situated near Geraldton in Western Australia \cite{SKA_redbook2020}.
SKA1-MID operates over the frequency range $350$--$1750~\mathrm{MHz}$, while SKA1-LOW covers $50$--$350~\mathrm{MHz}$, allowing the SKA to probe large-scale structure over a wide range of redshifts using the redshifted 21-cm line. In this work, we focus on SKA1-MID, as its frequency coverage corresponds to the redshift range $z \lesssim 3$, which is particularly relevant for studies of late-time cosmology and the dark sector \cite{Chen2020}. Only the single-dish (auto-correlation) mode is considered here, while the interferometric mode is not included in the our work.

The SKA1-MID instrument will operate primarily in two frequency bands of cosmological interest. Band~1 covers the frequency interval $350$--$1050~\mathrm{MHz}$, corresponding to a redshift range $0.35 < z < 3.06$, while Band~2 spans $950$--$1750~\mathrm{MHz}$, probing the low-redshift Universe over $0 < z < 0.49$ via the 21-cm emission line \cite{SKA_redbook2020}. SKA1-MID is a dish array composed of multiple sub-arrays, including the 64 dishes of the MeerKAT precursor instrument, each with a diameter of $13.5~\mathrm{m}$, and 133 newly built SKA1 dishes with diameters of $15~\mathrm{m}$. Following the treatment adopted in \cite{Chen2020}, we assume that all 197 dishes can be approximated as having a uniform diameter of $15~\mathrm{m}$ and are equipped with dual-polarization receivers, an assumption that does not significantly affect the resulting cosmological forecasts. The beam full-width at half-maximum is evaluated at the median frequency of each band. For Band~1, the angular resolution is $\theta_{\rm FWHM}=1.96^\circ$ at a median frequency $\nu_{\rm med}=700~\mathrm{MHz}$, while for Band~2 it is $\theta_{\rm FWHM}=1.17^\circ$ at $\nu_{\rm med}=1177.5~\mathrm{MHz}$ \cite{Chen2020}. For consistency and to facilitate a direct comparison with the BINGO experiment, we adopt a frequency channel width of $10~\mathrm{MHz}$ for both SKA1-MID bands throughout this work.

Following \cite{SKA_redbook2020}, we can calculate the system temperature of SKA1-MID array through
\begin{equation}
    T_{{\rm sys}}=T_{{\rm rx}} + T_{{\rm spl}} + T_{{\rm CMB}} + T_{{\rm gal}} ,
\label{eq:system temperature}
\end{equation}
where $T_{{\rm spl}}\approx 3 \rm K$ contributes from  ‘spill-over’, and the CMB temperature is $T_{{\rm CMB}}\approx 2.73\rm K$. $T_{{\rm gal}}$ contributes from our own galaxy as a function of frequency, as given by 
\begin{equation}
    T_{{\rm gal}}=25{\rm K} (408{\rm MHz}/\nu)^{2.75}.
\label{eq:galaxy}
\end{equation}
$T_{{\rm rx}}$ is the receiver noise temperature, for Band 1, $T_{{\rm rx}}=15{\rm K}+30{\rm K}(\frac{\nu}{{\rm GHz}}-0.75)^2$; for Band 2, $T_{{\rm rx}}=7.5 \rm K$.  For Band 2, we can assume a constant Galactic contribution of $T_{{\rm gal}}\approx 1.3\rm K$, since at high frequencies of the operation frequency range of Band 2, it is subdominant. So, the system temperature for Band 2 is constant, $T_{{\rm sys}}=15\rm K$.  And the system temperature for BINGO is also a constant, $70\rm K$. The sky coverage for SKA1-MID Band 1 and Band 2 are assumed to be 20000 and 5000 ${\rm deg}^2$ respectively, and assuming one year operation time. Details of the configurations of BINGO and SKA1-MID are summarized in Table \ref{tab:surveys}.

\subsection{noises}

In this work we include the thermal noise and the shot noise. The thermal noise is from the voltages generated by thermal agitations in the resistive components of the receiver. Then, the pixel noise is given by 
\begin{equation}
    \sigma_{\rm T}=\frac{T_{{\rm sys}}}{\sqrt{t_{{\rm pix}}\delta\nu}},
\label{eq:pixel noise}
\end{equation}
where $\delta\nu$ is the channel bandwidth, $T_{{\rm sys}}$ is the total system temperature, $t_{{\rm pix}}$ is the integration time per pixel, obtained from
\begin{equation}
    t_{{\rm pix}}=t_{{\rm obs}}\frac{n_{{\rm beam}}n_{\rm d}\Omega_{{\rm pix}}}{\Omega_{{\rm sur}}} . 
\end{equation}
Where $t_{{\rm obs}}$ is the opetaion time of one year of the survey, $n_{{\rm beam}}$ and $n_{\rm d}$ is the number of beams and dishes, respectively. $\Omega_{{\rm pix}}\propto \theta_{{\rm FWHM}}^2$ is the sky area per pixel, and $\Omega_{{\rm sur}}$ is the total sky area of the survey. Then the angular power spectrum of thermal noise is
\begin{equation}
    N_l(z_i,z_j)=\frac{4\pi}{N_{{\rm pix}}}\sigma_{{\rm T},i} \sigma_{{\rm T},j}.
\label{eq:thermal}
\end{equation}
where $N_{{\rm pix}}$ is the number of pixels in the map, $\sigma_{{\rm T},i},\sigma_{{\rm T},j}$ are that in the equation (\ref{eq:pixel noise}) at different redshift.  Here, we assume $N_l(z_i,z_j)=0$ if $z_i\neq z_j$.  

Equation (\ref{eq:thermal}) is for a perfect beam receiver measuring, when considering the beam resolution $\theta_{{\rm FWHM}}$, there is beam corrections to the angular power spectrum of the thermal noise, reducing the signal by a factor of $b_l^2$ \cite{Chen2020},
\begin{equation}
    b_l(z_i)={\rm exp}(-\frac{1}{2}l^2\sigma_{b,i}^2),
\end{equation}
where $\sigma_{b,i}=\theta_{\rm B}(z_i)/\sqrt{8{\rm ln}2}$ \cite{Bull2015} and $\theta{\rm _B}(z_i)=\theta_{{\rm FWHM}}(\nu_{{\rm med}})\nu_{{\rm med}}/\nu_i$.

The thermal noise is from the instrument, while the shot noise is actually a part of the signal itself. Since the HI sources are discrete, it will contribute to the auto-spectra measured in the signal, behave as $C_l^{{\rm shot}}=\bar T_{\rm b}^2(z)/\bar N(z)$ \cite{Hall2013}. Recall that  $\bar T_{\rm b}(z)$ is the average brightness temperature of the 21-cm emission line, and $\bar N(z)$ is the angular density of the sources. $\bar N(z)$ is given by 
\begin{equation}
    \bar N(z)=\frac{n_0 c}{H_0}\int\frac{\chi^2(z)}{E(z)}{\rm d}z ,
\end{equation}
 where $n_0=0.03h^3\rm Mpc^{-3}$ \cite{Masui2010} is the comoving number density of sources.

Foreground contamination is another important systematic for HI intensity mapping. At $\nu\sim 1\,{\rm GHz}$, Galactic synchrotron emission and unresolved extragalactic point sources are typically about $10^{4}$ brighter than the HI signal, so an efficient foreground-cleaning procedure is required before cosmological inference. Following the discussion in Sec.~3.3 of \cite{Xiao2021}, in the baseline forecast we adopt an optimistic assumption that the dominant smooth-spectrum foregrounds can be removed and only subdominant residuals remain; their impact can be incorporated as an additional covariance contribution and is expected to mainly affect the largest angular scales. See also \cite{Bigot-Sazy:2015jaa} and \cite{Olivari:2015tka} for foreground subtraction in HI intensity mapping.

\begin{figure}[H]
     \centering
     \includegraphics[width=0.9\linewidth]{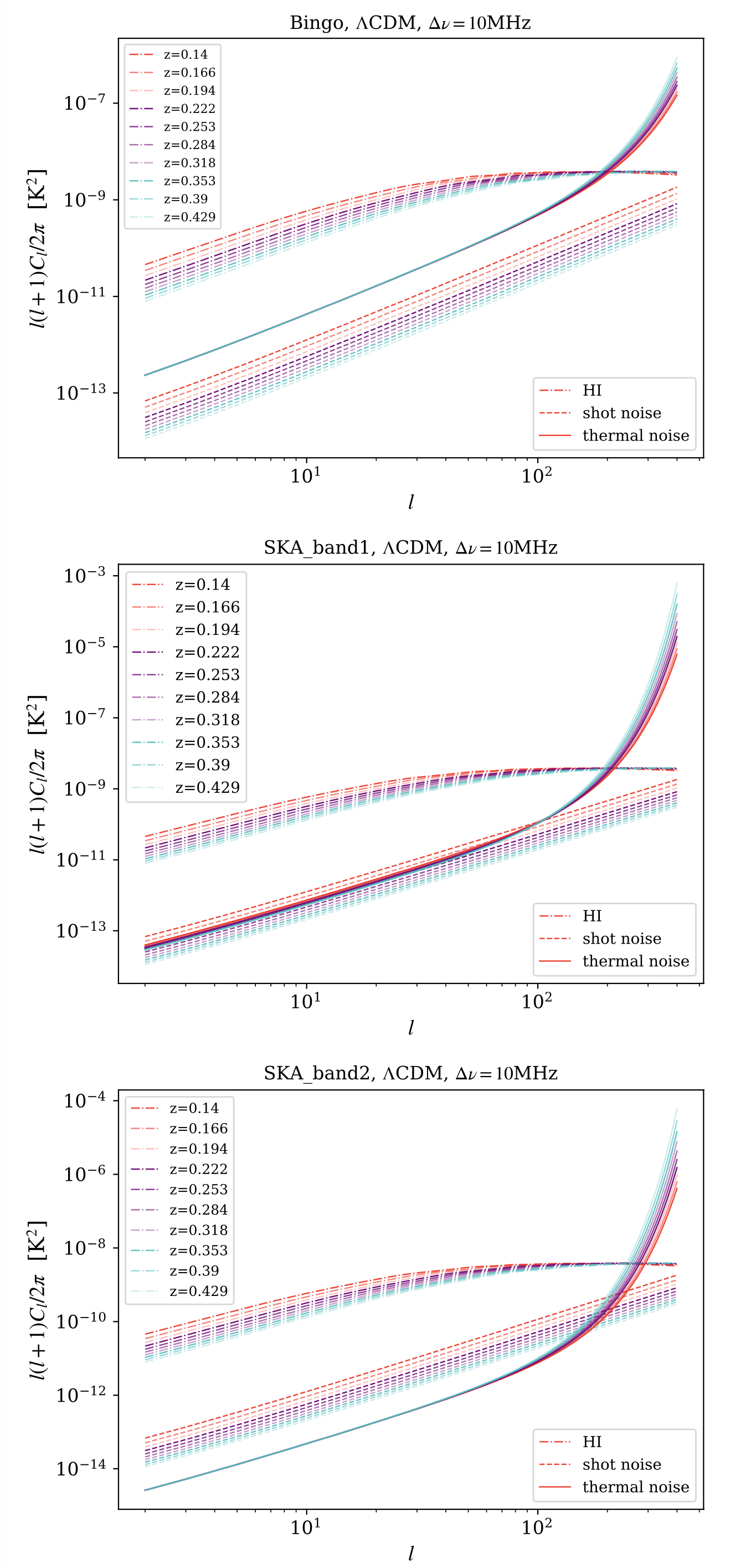}
    \caption{Angular power spectra of the HI 21-cm signal, thermal noise, and shot noise for BINGO (top), SKA1-MID Band~1 (middle), and Band~2 (bottom), evaluated at multiple redshift bins. The HI signal dominates over instrumental noise on large angular scales. The thermal noise contribution is smallest for SKA Band~2.}     
     \label{fig:surveys}
\end{figure}

In Figure \ref{fig:surveys}, we plot the auto-spectra for the HI signal, the thermal noise and shot noise. The top left panel shows the plot for BINGO at ten different redshifts, with the channel bandwidth being 10 MHz.  It shows that the auto-spectra for HI signal is larger than that of the thermal noise, and the shot noise is the smallest.  For all of the three kinds of auto-spectra, as the redshift goes higher, the auto-spectra becomes smaller. The top right and bottom panels show the plots for SKA1-MID Band 1 and 2, respectively. Since the HI signal and shot noise auto-spectra are identical across BINGO, SKA1-MID Band 1, and SKA1-MID Band 2, the thermal noise auto-spectra can be directly compared: SKA1-MID Band 2 exhibits the lowest amplitude, followed by SKA1-MID Band 1, while BINGO yields the highest thermal noise auto-spectrum.

 \subsection{Fisher matrix}

We use Fisher matrix to estimate the uncertainty of the parameters from the IM data. For a set of cosmological parameters 
\begin{equation}
    \boldsymbol{\theta}=\{\Omega_bh^2,  \Omega_ch^2,{\rm ln}(10^{10}A_s),n_s,h, B_0,b_{{\rm HI}}, \Omega_{\rm HI}\!\cdot\! b_{\rm HI}\} \},
\end{equation}
the Fisher matrix for the parameters $\theta_i$ is defined as the ensemble average of the Hessian matrix of the log-likelihood function \cite{Asorey2012}
\begin{equation}
    F_{ij}=\Big < -\frac{\partial{\rm ln}\mathcal L}{\partial\theta_i \partial\theta_j} \Big > =\frac{1}{2}{\rm Tr}\Big [ \textbf{\textit{C}}^{-1} \frac{\partial \textbf{\textit{C}}}{\partial\theta_i} \textbf{\textit{C}}^{-1} \frac{\partial \textbf{\textit{C}}}{\partial\theta_j} \Big].
\end{equation}
Here the covariance matrix $\textbf{\textit{C}}$ is given by
\begin{equation}
    \textbf{\textit{C}} =\left [  \begin{array}{ccccc}
    A_{l=2} & 0 & ... & 0 & \\
    0 & A_3 & ... & 0 & \\
    \vdots & \vdots & ... &\vdots &\\
    0 & 0 & ...& A_n & \\
\end{array}
 \right ] ,
\end{equation}
where 
\begin{equation}
    A_l=(2l+1)\left [ \begin{array}{ccccc}
    C_l(z_1,z_1) & C_l(z_1,z_2) & ... & C_l(z_1,z_n) &\\
    C_l(z_2,z_1) & C_l(z_2,z_2) & ... & C_l(z_2,z_n) &\\
    \vdots & \vdots & ... &\vdots &\\
    C_l(z_n,z_1) & C_l(z_n,z_2) & ... & C_l(z_n,z_n) &\\
\end{array}
\right]
\end{equation}
and 
\begin{equation}
    C_l(z_i,z_j)=C_l^{{\rm HI}}(z_i,z_j)+\delta_{ij}C_l^{{\rm shot}}(z_i,z_j)+N_l(z_i,z_j)B_l(z_i,z_j).
\end{equation}

\section{Forecast Results}
\label{sec:results}

\begin{figure}[H]
     \centering
     \includegraphics[width=0.7\linewidth]{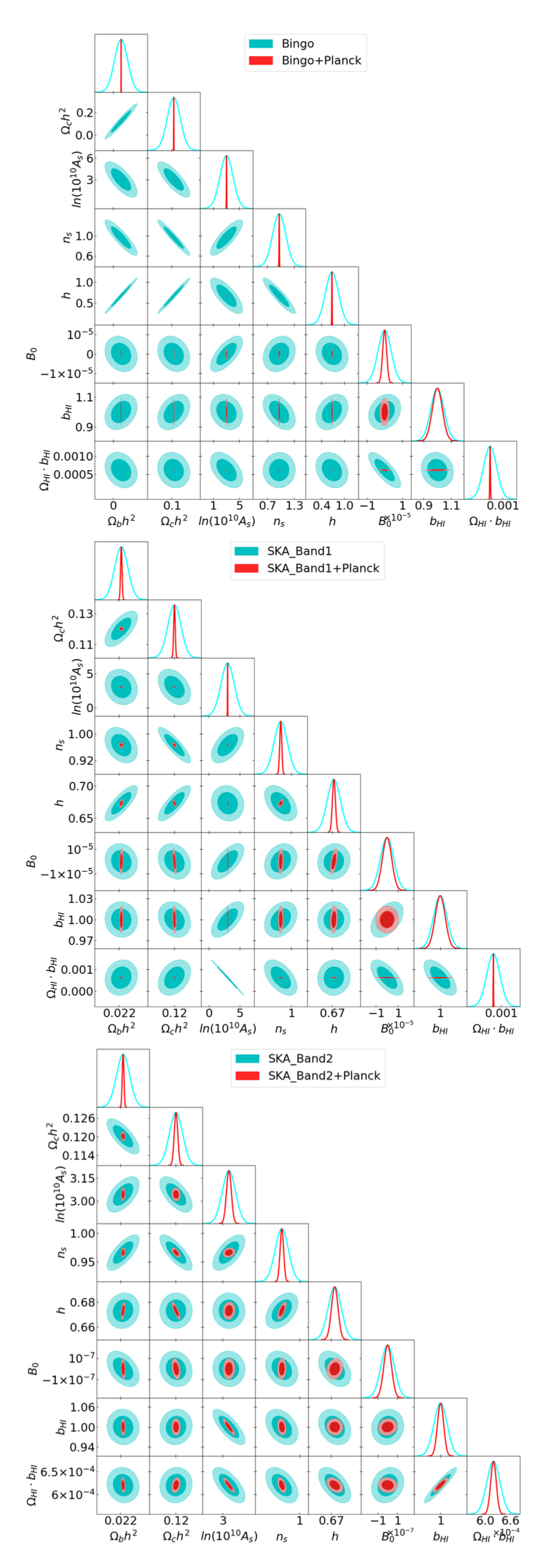}
     \caption{Marginalized one- and two-dimensional confidence regions (68\% and 95\%) for selected cosmological parameters and the $f(R)$ gravity parameter $B_0$. Results are shown for BINGO (top), SKA1-MID Band~1 (middle), and Band~2 (bottom), both with and without Planck CMB priors. The inclusion of Planck data significantly reduces parameter degeneracies and tightens the constraints.}
     \label{fig:triangle}
 \end{figure}

\begin{figure}[H]
     \centering
     \includegraphics[width=0.9\linewidth]{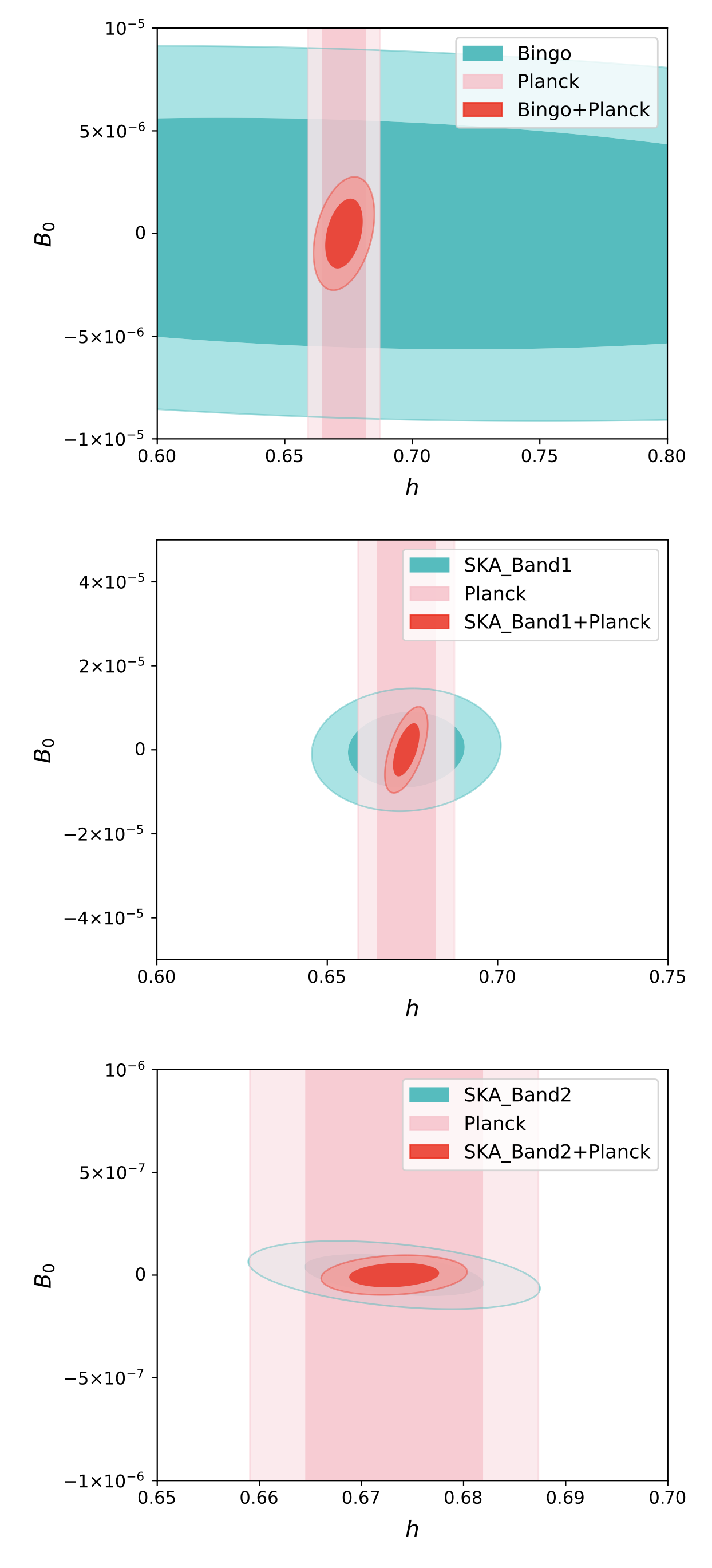}
     \caption{Marginalized 68\% and 95\% confidence contours in the $B_0$–$h$ plane for BINGO (top), SKA1-MID Band~1 (middle), and Band~2 (bottom). Results are shown for IM-only forecasts and in combination with Planck priors. The strong degeneracy between $B_0$ and $h$ present in IM-only surveys is efficiently broken when CMB information is included.}
     \label{fig:b0h}
 \end{figure}

In this section, we present Fisher matrix forecasts for constraining cosmological parameters and the $f(R)$ gravity parameter $B_0$ using upcoming HI 21-cm IM surveys. We consider three experimental configurations, BINGO, SKA1-MID Band 1, and SKA1-MID Band 2. And we further quantify the gain in constraining power achieved by combining each IM survey with Planck CMB priors. The marginalized $1\sigma$ uncertainties on all parameters are summarized in Table \ref{tab:constraints}, and the key parameter degeneracies are illustrated in Figures \ref{fig:triangle} and \ref{fig:b0h}.

We forecast the marginalized constraints on the parameters $\boldsymbol{\theta}=\{\Omega_b h^2,\,\Omega_c h^2,\,{\rm ln}(10^{10}A_s),\,n_s,\,h,\,B_0,\,b_{\rm HI},\, \Omega_{\rm HI}\!\cdot\! b_{\rm HI}\}$, around a fiducial $\Lambda$CDM background with $B_0=0$ (see Table~\ref{tab:constraints} for the fiducial values). As expected, BINGO alone yields relatively weak constraints on the standard cosmological parameters, primarily due to its limited sky coverage and narrow redshift range. Remarkably, however, even this pathfinder IM experiment exhibits measurable sensitivity late-time modifications of gravity, yielding  $\sigma(B_0)\simeq 3.73\times 10^{-6}$. This result confirms that low-redshift 21-cm IM surveys can probe $f(R)$ gravity not only via background evolution but, more robustly, through its distinct imprint on the growth of cosmic structure. SKA1-MID shows substantially stronger constraining power. 
In particular, although SKA Band 2 has a smaller survey area than Band 1, it gives a much tighter constraint on \(B_0\). This is mainly because Band 2 probes lower redshifts, where the late-time effects of \(f(R)\) gravity are stronger, and because its lower system temperature leads to smaller instrumental noise.

With its larger sky coverage, broader redshift range and lower instrumental noise, both Band 1 and Band 2 of SKA1-MID reduce the marginalized $1\sigma$ uncertainties on $(\Omega_b h^2,\Omega_c h^2,A_s,n_s,h)$ by approximately an order of magnitude  relative to BINGO. In particular, SKA Band 2 achieves  $\sigma(h)\simeq 5.84\times 10^{-3}$, $\sigma(B_0)\simeq 6.78\times 10^{-8}$, highlighting its excellent sensitivity to scale-dependent growth signatures induced by modified gravity at low redshift.  We also find that the $b_{\rm HI}$ and $\Omega_{\rm HI}\!\cdot\! b_{\rm HI}$ are constrained at the levels of $\sigma(b_{\rm HI})\simeq2.19\times10^{-2}$ and $\sigma(\Omega_{\rm HI}\!\cdot\! b_{\rm HI})\simeq1.59\times10^{-5}$ for SKA Band~2, while the corresponding BINGO constraints are significantly weaker.
When Planck priors are included, the constraints on the parameters $(\Omega_b h^2,\Omega_c h^2,A_s,n_s,h)$ tighten significantly for all IM surveys, demonstrating the powerful complementarity between early-universe CMB probes and late-time large-scale structure measurements. Importantly, IM data also break key degeneracies affecting $B_0$, leading to a marked improvement in its determination over Planck alone. For example, combining BINGO with Planck yields  $\sigma(B_0)\simeq 1.13\times 10^{-6}$, while the SKA Band 2 and Planck combination achieves the strongest constraint to date,  $\sigma(B_0)\simeq 3.95\times 10^{-8}$. 
 In the joint analyses, the constraints on $b_{\rm HI}$ and $\Omega_{\rm HI}\!\cdot\! b_{\rm HI}$ are also improved, with SKA Band~2 + Planck giving $\sigma(b_{\rm HI})\simeq1.03\times10^{-2}$ and $\sigma(\Omega_{\rm HI}\!\cdot\! b_{\rm HI})\simeq6.78\times10^{-6}$.

Figure \ref{fig:triangle} shows the marginalized two-dimensional confidence contours (68\% and 95\%). For IM-only forecasts, strong degeneracies appear between $B_0$ and late-time parameters that affect the amplitude and evolution of clustering, especially the Hubble parameter $h$, the HI bias $b_{\rm HI}$,  and $\Omega_{\rm HI}\!\cdot\! b_{\rm HI}$. This is expected because modified gravity can partially mimic changes in the late-time expansion rate and growth history. These degeneracies are efficiently reduced once Planck priors are included, since CMB data strongly constrain the early-Universe background cosmology and primordial parameters. The joint IM+Planck constraints therefore isolate the late-time, scale-dependent signatures characteristic of $f(R)$ gravity, leading to significantly smaller confidence regions.

The degeneracy between $B_0$ and $h$ is shown explicitly in Figure \ref{fig:b0h}, where we plot the 68\% and 95\% confidence contours for BINGO, SKA Band 1, and SKA Band 2, both alone and in combination with Planck. For IM-only surveys, the contours are elongated in the $B_0$--$h$ plane, indicating a strong degeneracy between these two parameters, with the effect being most pronounced for BINGO and reduced for SKA configurations due to their improved sensitivity. After adding Planck priors, the contours shrink substantially and become much tighter in $h$, showing that the uncertainty in the expansion history no longer dominates the determination of $B_0$. Among all combinations, SKA Band~2 + Planck provides the tightest constraints, consistent with the smallest $\sigma(B_0)$ reported in Table~\ref{tab:constraints}.

It is useful to interpret these forecasted constraints in the broader context of current limits from optical large-scale-structure probes. Recent galaxy-clustering, weak-lensing, and cluster-abundance analyses already provide stringent tests of modified gravity, and in some cases lead to similar bounds than those forecast here \cite{Ishak:2024jhs,Artis:2024eco,Vogt:2024SPTfR, Bai:2024hpw}. Therefore, the value of HI intensity mapping lies not only in its competitive sensitivity to $B_0$, but also in its role as an independent probe with different observational systematics. In particular, future combined analyses of HI intensity mapping with optical surveys and CMB data are expected to further break parameter degeneracies and improve the robustness of the final constraints on $B_0$.

\begin{figure}[H]
     \centering
      \includegraphics[width=1.\linewidth]{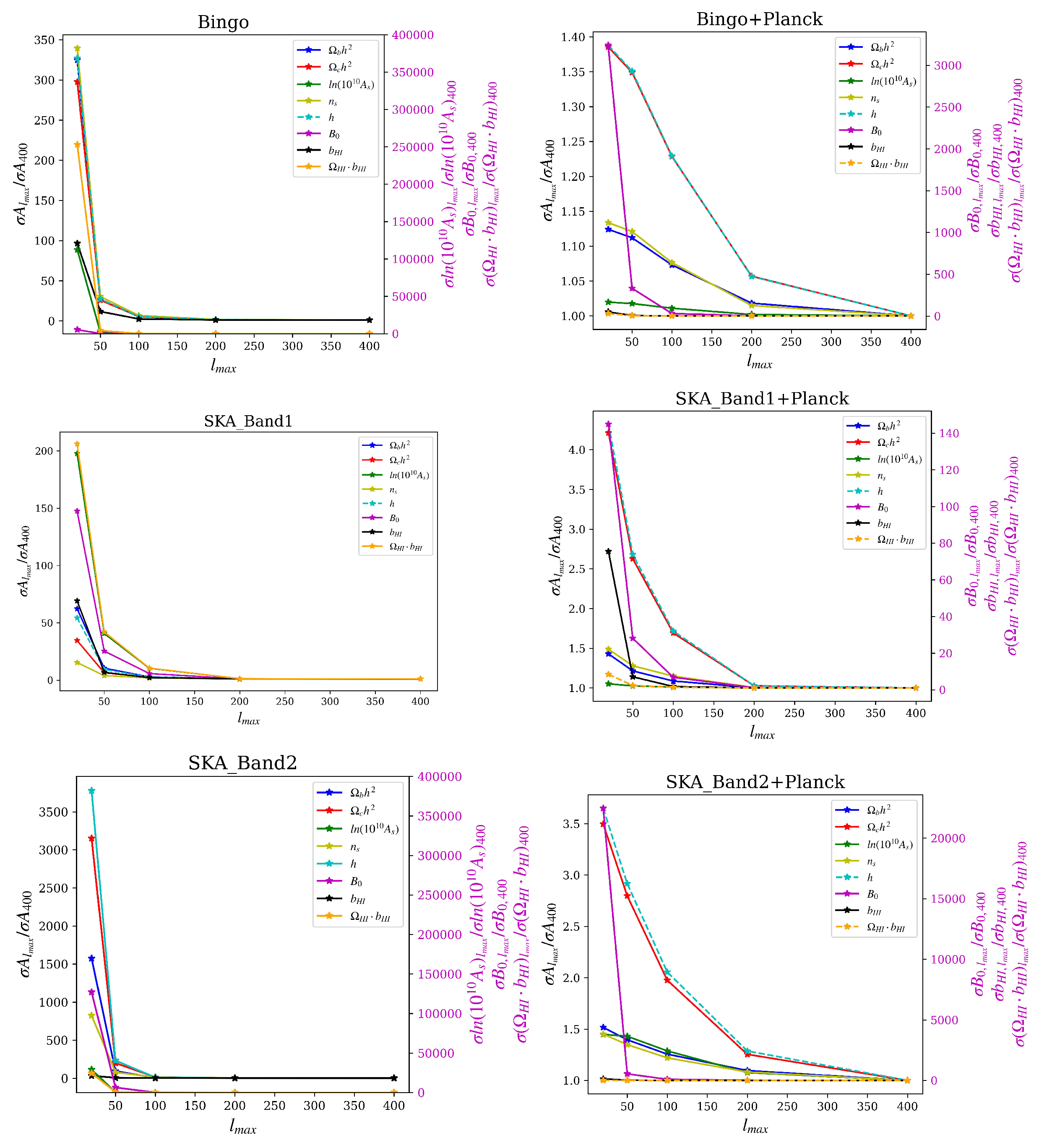}
     \caption{Ratios of the marginalized parameter uncertainties obtained with a maximum multipole $\ell_{\rm max}$ to those obtained with $\ell_{\rm max}=400$, as functions of $\ell_{\rm max}$. Results are shown for different survey configurations. A dual $y$-axis is adopted to display parameters with different amplitudes of $\sigma(\theta_i)/\sigma(\theta_i;\ell_{\rm max}=400)$. The figure illustrates the convergence of parameter constraints as smaller angular scales are progressively included, showing that most constraints saturate at $\ell_{\rm max}\sim300$--$400$.}
     \label{fig:lmax}
 \end{figure}

Figure \ref{fig:lmax} shows the ratio of the marginalized uncertainties of cosmological parameters obtained by truncating the angular power spectrum at a given maximum multipole $\ell_{\rm max}$ to those obtained with $\ell_{\rm max}=400$, i.e. $\sigma(\theta_i;\,\ell_{\rm max})/\sigma(\theta_i;\,\ell_{\rm max}=400)$,
as a function of $\ell_{\rm max}$. This definition of the vertical axis follows that adopted in Figure~11 of Ref.~\cite{Xiao2021}, allowing a direct assessment of how the parameter constraints converge as smaller angular scales are progressively included. For all survey configurations, the ratios decrease rapidly with increasing $\ell_{\rm max}$ and approach unity at $\ell_{\rm max}\simeq 300$--$400$, indicating that the constraints on most parameters are already close to saturation once multipoles up to a few hundred are included. This behavior demonstrates that the dominant Fisher information is carried by large- and intermediate-scale modes, while the contribution from very high multipoles is subdominant. For clarity, Figure~\ref{fig:lmax} is plotted using a dual $y$-axis, where parameters with significantly different amplitudes of $\sigma(\theta_i)/\sigma(\theta_i;\ell_{\rm max}=400)$ are separated onto the left and right axes. This presentation highlights the relative convergence rates of different parameters without affecting the physical interpretation. In particular, late-time parameters such as $B_0$, $b_{\rm HI}$ and $\Omega_{\rm HI}\cdot b_{\rm HI}$ exhibit a slower convergence with $\ell_{\rm max}$ than background parameters, reflecting their stronger sensitivity to large-scale modes. This explains why the constraints on $B_0$ are primarily driven by low-$\ell$ information, consistent with the results shown in Figures. \ref{fig:triangle} and \ref{fig:b0h}.

\begin{figure}[H]
     \centering
     \includegraphics[width=0.9\linewidth]{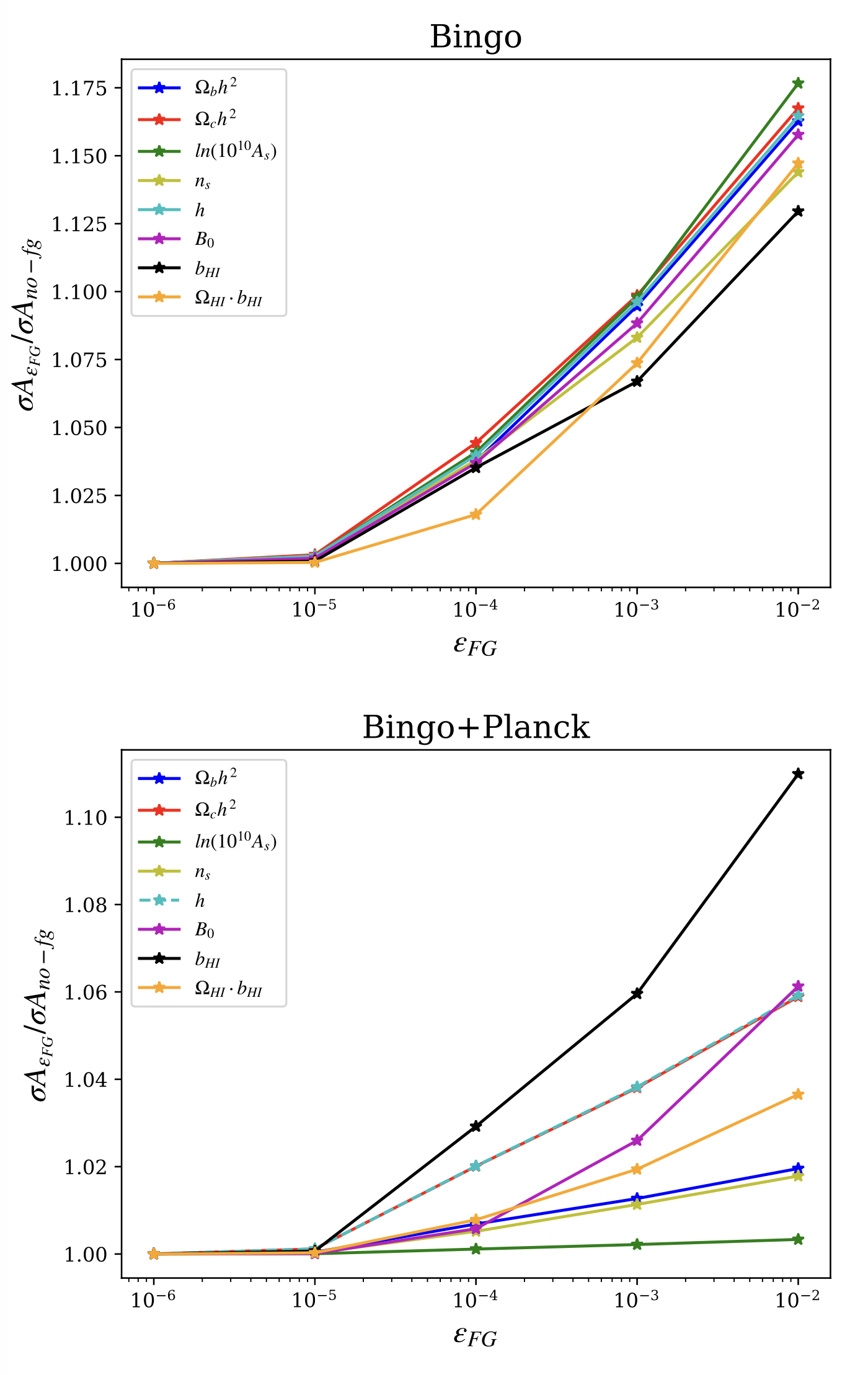}
     \caption{Ratio of marginalized parameter uncertainties obtained with foreground residuals to those obtained without foregrounds, $\sigma A_{\epsilon_{\rm FG}}/\sigma A_{\rm no\text{-}fg}$, as a function of the foreground residual amplitude $\epsilon_{\rm FG}$. Results are shown for BINGO, with and without Planck priors. As the foreground residual level increases, the parameter constraints are gradually degraded. The impact is more pronounced for late-time parameters, especially $B_0$, $b_{\rm HI}$, and $\Omega_{\rm HI} b_{\rm HI}$, while the inclusion of Planck priors reduces the relative degradation by constraining the background and primordial cosmological sectors.}
     \label{fig:foreground}
 \end{figure}

Figure \ref{fig:foreground} further quantifies the impact of foreground residuals on the parameter constraints by showing the ratio $\sigma A_{\epsilon_{\rm FG}}/\sigma A_{\rm no-fg}$ as a function of the foreground amplitude parameter $\epsilon_{\rm FG}$, where $\sigma A_{\epsilon_{\rm FG}}$ and $\sigma A_{\rm no-fg}$ denote the marginalized uncertainties obtained with and without the foreground residuals, respectively. Similar to the presentation in Figure~13 of Ref.~\cite{Xiao2021}, this figure directly illustrates the degradation of the Fisher constraints induced by residual foreground contamination. As expected, the parameter uncertainties increase monotonically as the foreground residual level becomes larger. For the cases shown here, the degradation remains moderate over the displayed range, indicating that the constraints are not immediately destroyed by small residual foregrounds, but the effect is nevertheless non-negligible, especially for the late-time parameters that are more directly tied to the amplitude and scale dependence of the HI clustering signal. After including Planck priors, the relative impact of foreground residuals is visibly reduced, because the CMB information already constrains the background and primordial sectors very efficiently. Therefore, Figure \ref{fig:foreground} confirms that although foreground residuals weaken the constraining power of HI intensity mapping, the combination with Planck priors helps to stabilize the final constraints and preserve sensitivity to the modified-gravity parameter $B_0$.

\begin{table*}[t]
\footnotesize
\tabcolsep 7pt 
\begin{tabular*}{\textwidth}{l|c|c|c|c|c|c|c|c}
\toprule
\hline
 & $\Omega_b h^2 $  & $\Omega_c h^2 $ &  ${\rm ln}(10^{10}A_s)$  & $n_s$  & $h$   & $B_0$  & $b_{HI}$ & $ \Omega_{HI}\cdot b_{HI}$  \\
  \hline 
Fiducial values  & 0.022383   & 0.12011   &  $3.044$  &  0.96605  &  0.6732  &   0.00  &   1.00 & 0.00062 \\
\hline 
Bingo &  $1.83 \times 10^{-2}$ &   $6.43\times 10^{-2}$  &  $1.01\times 10^{0}$  &  $1.46\times 10^{-1}$ &  $1.73\times 10^{-1}$ &  $3.73\times 10^{-6}$ &  $4.81\times 10^{-2}$ & $2.00\times 10^{-4}$\\
SKA\_Band1 &  $1.10\times 10^{-3}$ &   $4.73\times 10^{-3}$  &  $1.08\times 10^{0}$  &  $2.17\times 10^{-2}$ &  $1.13\times 10^{-2}$ &  $5.98\times 10^{-6}$ &  $1.03\times 10^{-2}$ &  $3.29\times 10^{-4}$\\
SKA\_Band2 &  $9.41\times 10^{-4}$ &   $2.33\times 10^{-3}$  &  $4.78\times 10^{-2}$  &  $1.27\times 10^{-2}$ &  $5.84\times 10^{-3}$ &  $6.78\times 10^{-8}$ &  $2.19\times 10^{-2}$&  $1.59\times 10^{-5}$ \\
Planck &  $1.53\times 10^{-4}$ &  $1.23\times 10^{-3}$  &  $2.16\times 10^{-2}$  &  $4.24\times 10^{-3}$  &  $5.87\times 10^{-3}$   &  $5.31\times 10^{-2}$  & - & - \\
Bingo+Planck &  $1.43\times 10^{-4}$ &   $1.03\times 10^{-3}$  &  $2.11\times 10^{-2}$  &  $3.92\times 10^{-3}$ &  $4.88\times 10^{-3}$ &  $1.13\times 10^{-6}$ &  $3.67\times 10^{-2}$ &  $9.30\times 10^{-6}$ \\
SKA\_Band1+Planck &  $1.23\times 10^{-4}$ &   $5.49\times 10^{-4}$  &  $2.07\times 10^{-2}$  &  $3.34\times 10^{-3}$ &  $2.57\times 10^{-3}$ &  $4.19\times 10^{-6}$ &  $7.76\times 10^{-3}$ &   $7.91\times 10^{-6}$ \\
SKA\_Band2+Planck &  $1.21\times 10^{-4}$ &  $ 6.29\times 10^{-4}$  &  $1.77\times 10^{-2}$  &  $3.43\times 10^{-3}$ &  $2.92\times 10^{-3}$ &  $3.95\times 10^{-8}$ &  $1.03\times 10^{-2}$&  $6.78\times 10^{-6}$ \\
\hline
\bottomrule
\end{tabular*}
\caption{Marginalized $1\sigma$ uncertainties on cosmological parameters and the $f(R)$ gravity parameter $B_0$ obtained from Fisher-matrix forecasts. Results are shown for BINGO, SKA1-MID Band~1, and Band~2, both alone and in combination with Planck CMB priors.}
\label{tab:constraints}
\end{table*}

\section{ Impact of redshift-dependent \texorpdfstring{$\Omega_{\rm HI}(z)$}{OmegaHI(z)} and \texorpdfstring{$b_{\rm HI}(z)$}{bHI(z)}}
\label{sec:results2}

In this section, we present a second set of Fisher-matrix forecasts for constraining cosmological parameters and the $f(R)$ gravity parameter $B_0$ using upcoming HI 21-cm intensity mapping surveys. In contrast to Section \ref{sec:results}, where the HI sector was parameterized phenomenologically, here we adopt explicit redshift-dependent forms for the neutral hydrogen abundance and HI bias. More specifically, we model the post-reionization HI sector using the redshift evolution prescription motivated by the MeerKAT/SKA HI intensity mapping literature, following the forms associated with Eqs.~(2)--(3) of Ref. \cite{MeerKLASS:2017vgf} in the MeerKLASS-related modeling framework and the fiducial trends shown in Figure~1 of Ref.~\cite{Santos:2015gra}. In practice, we set the overall prefactors of these expressions to unity, so that $\Omega_{\rm HI}(z)$ and $b_{\rm HI}(z)$ contain no additional free parameters. The HI sector is therefore completely specified by redshift alone in this section. A commonly used realization of this prescription is
\begin{eqnarray}
\Omega_{\rm HI}(z)&=&4.8304\times 10^{-4}+3.8356\times 10^{-4}z- 6.5119\times 10^{-5}z^2, \\
b_{\rm HI}(z)&=&0.66655+0.17765z+0.050223z^2.
\end{eqnarray}
which is consistent with the fiducial HI evolution adopted in MeerKAT intensity mapping forecasts and with the trends shown in Figure~1 of Ref.~\cite{Santos:2015gra}. Consequently, the parameter set in this section is reduced to
\begin{equation}
\boldsymbol{\theta}=\{\Omega_b h^2,\ \Omega_c h^2,\ \ln(10^{10}A_s),\ n_s,\ h,\ B_0\},
\end{equation}
since neither $\Omega_{\rm HI}(z)$ nor $b_{\rm HI}(z)$ is treated as an independent nuisance parameter.

The marginalized $1\sigma$ uncertainties are summarized in Table \ref{tab:constraints2}. As expected, BINGO alone yields relatively weak constraints on the standard cosmological parameters because of its limited sky coverage and narrow redshift range, but it still exhibits sensitivity to modified gravity, with $\sigma(B_0)\simeq 3.77\times10^{-6}$. SKA1-MID provides significantly stronger constraining power. In the IM-only case, SKA Band~1 gives $\sigma(B_0)\simeq 4.16\times10^{-6}$, while SKA Band~2 reaches $\sigma(B_0)\simeq 1.14\times10^{-7}$, making it the most sensitive configuration among the three surveys. The corresponding uncertainty on the Hubble parameter is also smallest for SKA Band~2, with $\sigma(h)\simeq 1.94\times10^{-3}$, reflecting the high statistical power of this low-redshift survey configuration. Overall, Table \ref{tab:constraints2} shows that fixing the redshift evolution of $\Omega_{\rm HI}(z)$ and $b_{\rm HI}(z)$ leads to a cleaner extraction of cosmological information from the HI signal, since the overall HI amplitude is no longer marginalized over as an independent degree of freedom. 

When Planck priors are included, the constraints on the standard cosmological parameters tighten significantly for all three IM surveys, again demonstrating the complementarity between early-universe CMB data and late-time HI large-scale-structure measurements. The constraints on $B_0$ are also improved, although the level of improvement depends on the survey configuration. In particular, BINGO+Planck yields $\sigma(B_0)\simeq 1.48\times10^{-6}$, SKA Band~1+Planck gives $\sigma(B_0)\simeq 3.63\times10^{-6}$, and SKA Band~2+Planck reaches the tightest bound, $\sigma(B_0)\simeq 7.40\times10^{-8}$. Compared with the results in Section \ref{sec:results}, these constraints should be interpreted as forecasts obtained under a fixed redshift-evolution model for the HI abundance and bias, rather than after marginalization over free HI-sector amplitudes.

\begin{figure}[H]
     \centering
     \includegraphics[width=0.8\linewidth]{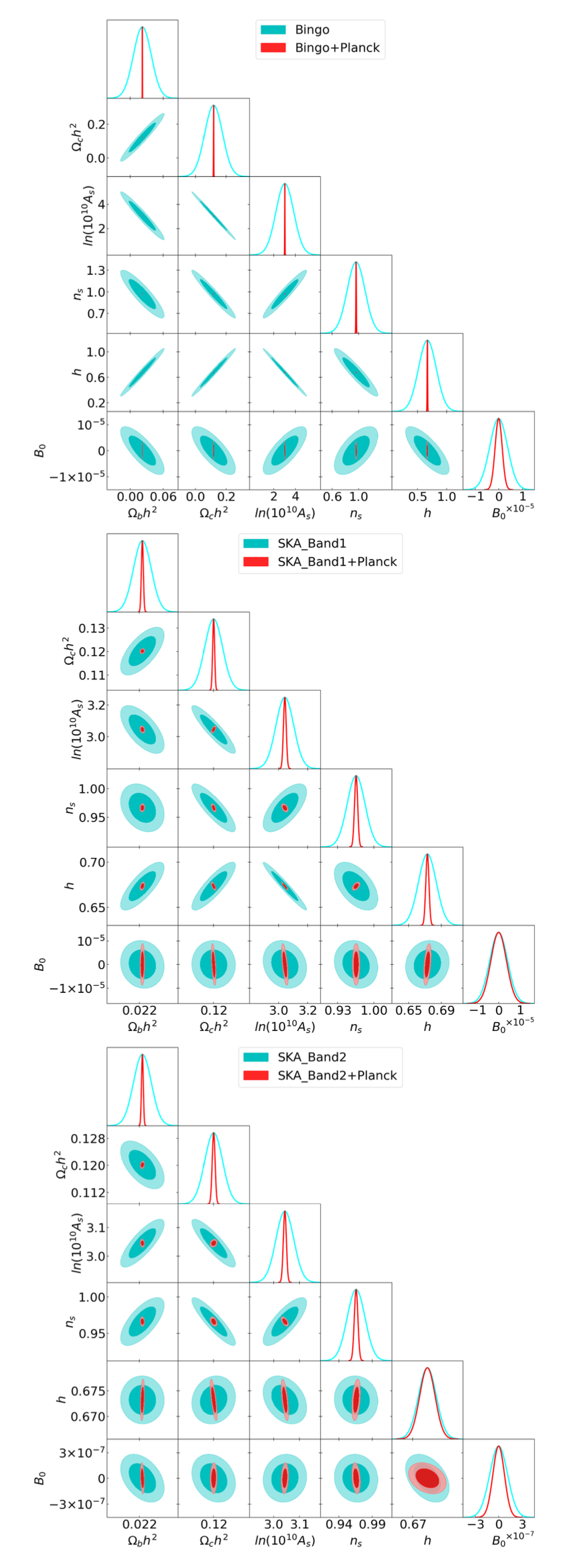}
     \caption{Same as Figure \ref{fig:triangle}, but for redshift-dependent \texorpdfstring{$\Omega_{\rm HI}(z)$}{OmegaHI(z)} and \texorpdfstring{$b_{\rm HI}(z)$}{bHI(z)}}
     \label{fig:triangle2}
 \end{figure}

Figure \ref{fig:triangle2} shows the marginalized one- and two-dimensional confidence regions for the selected cosmological parameters and $B_0$, for BINGO, SKA Band~1, and SKA Band~2, both with and without Planck priors. The overall structure of the contours is similar to that found in Section \ref{sec:results}, but the physical interpretation is different: here the HI signal is determined by the prescribed redshift evolution of $\Omega_{\rm HI}(z)$ and $b_{\rm HI}(z)$, with no additional nuisance freedom associated with constant HI amplitudes. As a result, the late-time parameter degeneracies are reduced relative to the more agnostic treatment adopted previously. In particular, the contours involving $B_0$ are generally tighter because the modified-gravity signal is no longer partially absorbed by free HI-amplitude parameters. As in Section \ref{sec:results}, the inclusion of Planck priors efficiently suppresses the remaining degeneracies by fixing the early-universe background and primordial sectors, thereby isolating the late-time, scale-dependent signatures characteristic of $f(R)$ gravity more clearly.

\begin{figure}[H]
     \centering
     \includegraphics[width=0.9\linewidth]{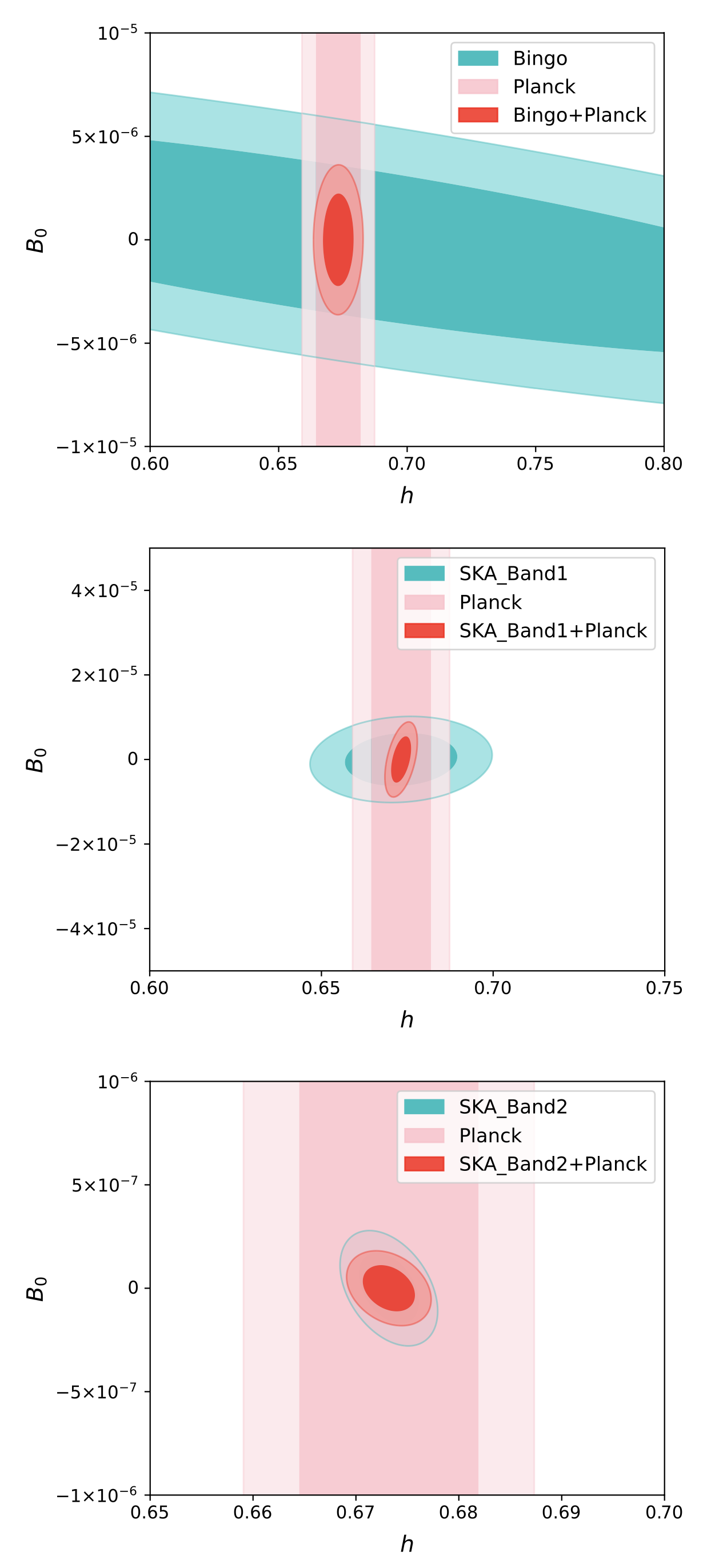}
     \caption{Same as Figure \ref{fig:b0h}, but for redshift-dependent \texorpdfstring{$\Omega_{\rm HI}(z)$}{OmegaHI(z)} and \texorpdfstring{$b_{\rm HI}(z)$}{bHI(z)}}
     \label{fig:b0h22}
 \end{figure}

The degeneracy between $B_0$ and $h$ is shown explicitly in Figure \ref{fig:b0h22}. For IM-only forecasts, the contours are still elongated in the $B_0$--$h$ plane, indicating that uncertainties in the late-time expansion history continue to affect the determination of $B_0$. However, once the redshift evolution of $\Omega_{\rm HI}(z)$ and $b_{\rm HI}(z)$ is fixed, this degeneracy is more directly associated with cosmology rather than with nuisance freedom in the HI sector. Among the three surveys, BINGO shows the broadest allowed region, SKA Band~1 yields a substantially tighter contour, and SKA Band~2 gives by far the smallest confidence region. After adding Planck priors, the contours contract markedly in the $h$ direction, showing that CMB information efficiently breaks the $B_0$--$h$ degeneracy. Consistent with Table \ref{tab:constraints2}, the combination SKA Band~2 + Planck provides the strongest overall constraint. This again reflects the fact that Band 2 is more sensitive to the low-redshift regime where the \(f(R)\) modification is more important, and also benefits from a lower noise level due to its smaller system temperature.

\begin{figure}[H]
     \centering
      \includegraphics[width=1.\linewidth]{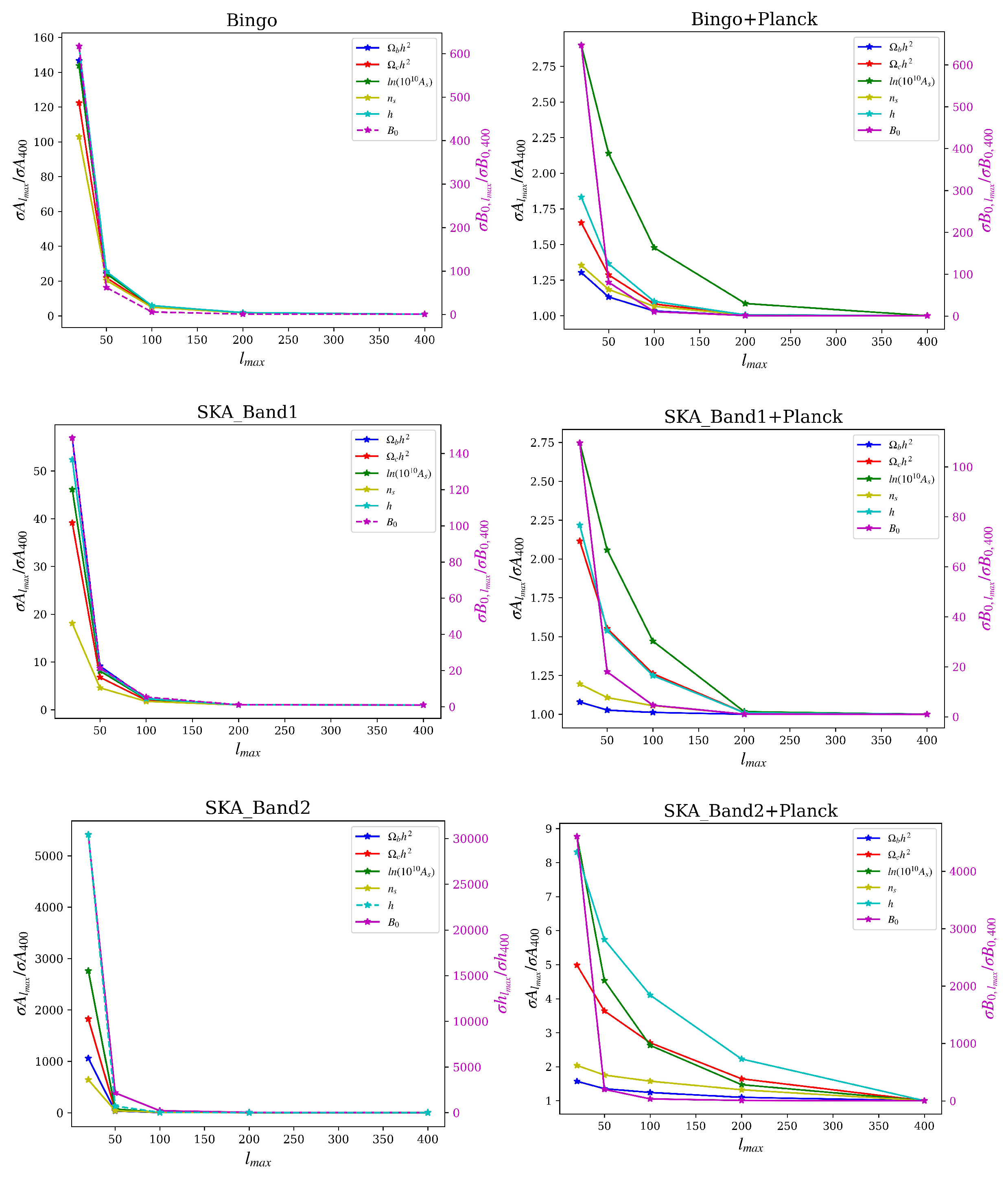}
     \caption{Same as Figure \ref{fig:lmax}, but for redshift-dependent \texorpdfstring{$\Omega_{\rm HI}(z)$}{OmegaHI(z)} and \texorpdfstring{$b_{\rm HI}(z)$}{bHI(z)}}
     \label{fig:lmax2}
 \end{figure}

Figure \ref{fig:lmax2} plays the same role in this section as Figure \ref{fig:lmax} does in Section \ref{sec:results} . It shows the ratios $\sigma(\theta_i;\ell_{\max})/\sigma(\theta_i;400)$ for the parameter set adopted here, namely $\theta=\{\Omega_b h^2,\Omega_c h^2,\ln(10^{10}A_s),n_s,h,B_0\}$, as functions of $\ell_{\max}$. Since the redshift evolution of $\Omega_{\rm HI}(z)$ and $b_{\rm HI}(z)$ is fixed by the fiducial model in this section, no additional HI-sector nuisance amplitudes are included in the Fisher analysis. As in Section \ref{sec:results}, the ratios decrease as $\ell_{\max}$ increases and gradually approach unity at $\ell_{\max}\sim 300$--$400$, indicating that most of the constraining power is already captured once multipoles up to a few hundred are included. This again shows that the sensitivity to the modified-gravity parameter $B_0$ is dominated mainly by large and intermediate angular scales, while the information gain from still higher multipoles is relatively limited. Compared with Figure \ref{fig:lmax}, the convergence pattern in Figure \ref{fig:lmax2} is cleaner and easier to interpret, because the HI sector is now completely specified by redshift-dependent functions rather than partially absorbed by free effective amplitudes. In particular, the convergence of the $B_0$ constraint remains slower than that of the standard background parameters, reflecting the fact that the signature of $f(R)$ gravity is primarily encoded in the late-time growth of structure. Therefore, Figure \ref{fig:lmax2} confirms that the main conclusions on the scale dependence of the $B_0$ constraint remain robust in the redshift-dependent HI model adopted in this section.

\begin{figure}[H]
     \centering
     \includegraphics[width=0.9\linewidth]{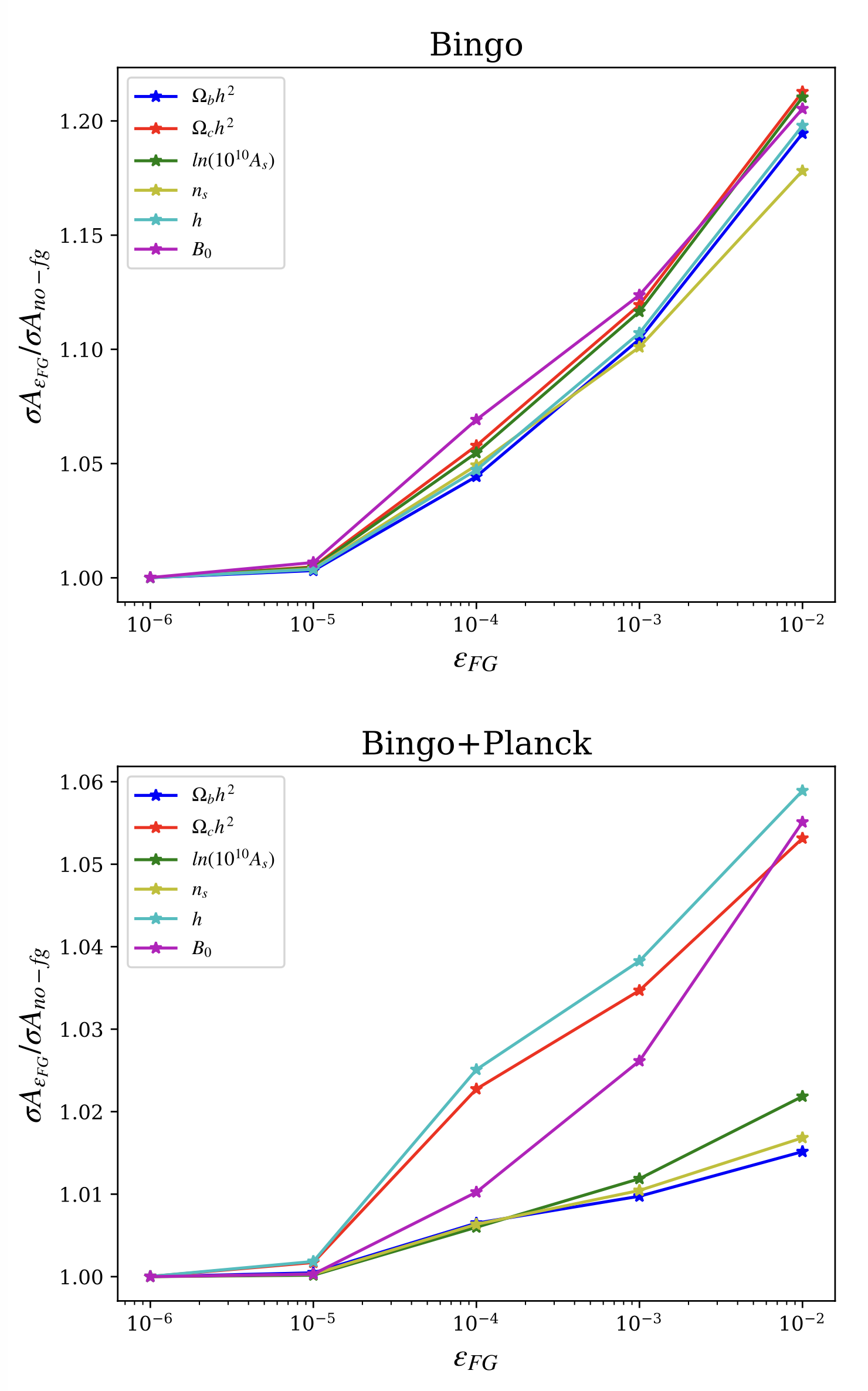}
     \caption{Same as Figure \ref{fig:foreground}, but for redshift-dependent \texorpdfstring{$\Omega_{\rm HI}(z)$}{OmegaHI(z)} and \texorpdfstring{$b_{\rm HI}(z)$}{bHI(z)}}
     \label{fig:foreground2}
 \end{figure}

Figure \ref{fig:foreground2} is the counterpart of Figure \ref{fig:foreground} in the redshift-dependent HI model considered in this section. It shows the ratio $\sigma A_{\epsilon_{\rm FG}}/\sigma A_{\rm no-fg}$ as a function of the foreground residual amplitude $\epsilon_{\rm FG}$, thereby illustrating how the forecasted parameter uncertainties are degraded once foreground contamination is taken into account. The qualitative behavior is the same as in Section \ref{sec:results} : larger residual foregrounds lead to systematically weaker constraints. However, because the HI sector is fixed here by the adopted fiducial evolution model for $\Omega_{\rm HI}(z)$ and $b_{\rm HI}(z)$, the foreground effect can be interpreted more directly as a loss of cosmological information rather than as an interplay with additional nuisance amplitudes. One finds that the degradation remains relatively controlled over the range shown, while the inclusion of Planck priors again reduces the relative sensitivity of the final constraints to foreground residuals. This means that, even in the redshift-dependent HI scenario, residual foregrounds should be accounted for carefully, but they do not qualitatively alter the conclusion that future HI intensity-mapping surveys, especially when combined with CMB priors, retain significant power to constrain the modified-gravity parameter $B_0$.

The main purpose of this section is therefore to present a complementary forecast in which the HI sector is modeled by fixed redshift-dependent functions rather than free effective amplitudes. In this setup, the constraints on $B_0$ reported in Table~3 and illustrated in Figs.~6 and 7 correspond specifically to the case where $\Omega_{\rm HI}(z)$ and $b_{\rm HI}(z)$ evolve with redshift according to the adopted fiducial model, with the prefactors fixed to unity and hence with no additional free HI parameters. This provides a useful benchmark for assessing how assumptions about the HI sector affect the inferred sensitivity of future HI intensity mapping surveys to modified gravity.

\begin{table*}[t]
\footnotesize
\tabcolsep 12pt 
\begin{tabular*}{\textwidth}{l|c|c|c|c|c|c}
\toprule
\hline
 & $\Omega_b h^2 $  & $\Omega_c h^2 $ &  ${\rm ln}(10^{10}A_s)$  & $n_s$  & $h$   & $B_0$    \\
  \hline 
Fiducial values  & 0.022383   & 0.12011   &  $3.044$  &  0.96605  &  0.6732  &   0.00   \\
\hline 
Bingo &  $1.61 \times 10^{-2}$ &   $5.80\times 10^{-2}$  &  $8.14\times 10^{-1}$  &  $1.36\times 10^{-1}$ &  $1.53\times 10^{-1}$ &  $3.77\times 10^{-6}$ \\
SKA\_Band1 &  $1.04\times 10^{-3}$ &   $4.15\times 10^{-3}$  &  $6.19\times 10^{-2}$  &  $1.72\times 10^{-2}$ &  $1.08\times 10^{-2}$ &  $4.16\times 10^{-6}$ \\
SKA\_Band2 &  $1.09\times 10^{-3}$ &   $2.88\times 10^{-3}$  &  $3.43\times 10^{-2}$  &  $1.34\times 10^{-2}$ &  $1.94\times 10^{-3}$ &  $1.14\times 10^{-7}$ \\
Planck &  $1.53\times 10^{-4}$ &  $1.23\times 10^{-3}$  &  $2.16\times 10^{-2}$  &  $4.24\times 10^{-3}$  &  $5.87\times 10^{-3}$   &  $5.31\times 10^{-2}$  \\
Bingo+Planck &  $1.27\times 10^{-4}$ &   $8.84\times 10^{-4}$  &  $1.18\times 10^{-2}$  &  $3.47\times 10^{-3}$ &  $3.95\times 10^{-3}$ &  $1.48\times 10^{-6}$ \\
SKA\_Band1+Planck &  $1.15\times 10^{-4}$ &   $4.62\times 10^{-4}$  &  $9.63\times 10^{-3}$  &  $3.06\times 10^{-3}$ &  $1.93\times 10^{-3}$ &  $3.63\times 10^{-6}$ \\
SKA\_Band2+Planck &  $1.10\times 10^{-4}$ &  $ 4.68\times 10^{-4}$  &  $5.84\times 10^{-3}$  &  $2.71\times 10^{-3}$ &  $1.68\times 10^{-3}$ &  $7.40\times 10^{-8}$ \\
\hline
\bottomrule
\end{tabular*}
\caption{Same as Tab. \ref{tab:constraints}, but for redshift-dependent \texorpdfstring{$\Omega_{\rm HI}(z)$}{OmegaHI(z)} and \texorpdfstring{$b_{\rm HI}(z)$}{bHI(z)}}
\label{tab:constraints2}
\end{table*}

\section{Conclusion and Discussion}
\label{sec:conclusions}

In this work, we have investigated the potential of future HI 21-cm intensity mapping surveys to constrain $f(R)$ gravity, focusing on the Compton wavelength parameter $B_0$. Using the Fisher matrix formalism, we forecast constraints from BINGO and SKA1-MID (Band~1 and Band~2), both alone and in combination with Planck CMB priors.

We find that low-redshift 21-cm IM surveys provide a powerful probe of modified gravity through their sensitivity to the growth of large-scale structure. Even BINGO alone is able to place meaningful constraints on $B_0$, while SKA1-MID significantly improves the precision due to its larger sky coverage and lower instrumental noise. Among the considered configurations, SKA Band~2 yields the tightest constraints, highlighting its strong sensitivity to scale-dependent growth effects induced by $f(R)$ gravity.

We further demonstrate that combining 21-cm IM data with Planck priors leads to substantial improvements in the constraints on $B_0$. The CMB data efficiently break degeneracies with background cosmological parameters, allowing the late-time information from 21-cm observations to isolate signatures of modified gravity. Our analysis shows that the dominant contribution to constraining $B_0$ arises from large-scale modes, consistent with the scale dependence of $f(R)$ effects.

Although HI intensity mapping can in principle provide the full 3-dimensional power spectrum (3D PS), in this work we adopt the tomographic angular power spectrum (APS) as a conservative and relativistically consistent probe. This formalism naturally incorporates projection effects and redshift-bin window functions, and avoids the flat-sky and distant-observer approximations. We stress that the angular power spectrum does not use the full line-of-sight information, so a future 3D analysis may further tighten the constraints on \(B_0\).

The APS provides a full-sky tomographic description and avoids the flat-sky approximation on large angular scales. However, Figure \ref{fig: cl_fR} shows that the fractional differences between different $B_0$ models at $\ell \lesssim 10$ have a similar trend to those at higher multipoles. Therefore, the largest-scale modes do not provide a qualitatively unique signature by themselves. Since the number of modes increases with $\ell$, a significant part of the constraining power comes from intermediate and smaller angular scales. In addition, wide-area APS analyses face additional foreground challenges, because Galactic foregrounds are highly anisotropic on large angular scales and foreground cleaning may remove or contaminate some low-$\ell$ modes. Thus, APS and 3D PS approaches are complementary.

Overall, our results indicate that upcoming HI intensity mapping experiments, especially SKA1-MID, will play a crucial role in testing deviations from General Relativity on cosmological scales.
Their strong constraining power comes not only from the large survey volume, but also from the continuous redshift coverage, the sensitivity to large-scale modes, and the reduced impact of shot noise compared with traditional galaxy surveys. In this sense, HI intensity mapping provides an important and complementary probe of \(f(R)\) gravity.
Finally, we emphasize that our forecasted constraints on $B_0$ should be interpreted in the broader context of existing limits from optical and other large-scale-structure probes. While such surveys already provide stringent tests of modified gravity, HI 21-cm intensity mapping offers an independent and largely systematics-orthogonal approach. Because HI intensity mapping, optical surveys, and CMB observations probe structure growth with different observational systematics and parameter degeneracies, their combination is expected to provide the most robust cosmological tests of $f(R)$ gravity in the future.

\Acknowledgements{We thank Chang Feng and Boyu Zhang for useful discussions. This work is supported by National Natural Science Fund of China under Grants Nos. 12175192, and Yangzhou Science and Technology Planning Project in Jiangsu Province of China (YZ2025233).}
\InterestConflict{The authors declare that they have no conflict of interest.}

\bibliographystyle{unsrt}
\bibliography{references}   

\end{multicols}
\end{document}